\documentclass[aps,pra,reprint,showkeys,amssymb,floatfix,longbibliography,superscriptaddress]{revtex4-2}

\usepackage{graphicx}
\usepackage{amsmath,amsfonts,amssymb}
\usepackage{hyperref}
\usepackage{braket}
\usepackage{xcolor}
\usepackage{tikz}
\usepackage{color}
\usepackage{enumerate}
\usepackage[caption=false]{subfig}
\usepackage{multirow}
\usepackage{qcircuit}
\usepackage[normalem]{ulem}
\usepackage{placeins}
\usepackage{xurl}
\usepackage{dsfont}
\usepackage{braket}
\usepackage[multidot]{grffile}
\usepackage{mathtools}
\usepackage{textcomp}
\newcommand{\unit}[1]{\ensuremath{\, \mathrm{#1}}}
\newcommand{\equref}[1]{Eq.~(\ref{#1})}

\begin{document}

\title{GPU-accelerated simulations of quantum annealing and the quantum approximate optimization algorithm}

\author{Dennis Willsch}
\thanks{Corresponding author: Dennis Willsch}
\email{d.willsch@fz-juelich.de}
\affiliation{Institute for Advanced Simulation, J\"ulich Supercomputing Centre,\\
Forschungszentrum J\"ulich, 52425 J\"ulich, Germany}
\author{Madita Willsch}
\affiliation{Institute for Advanced Simulation, J\"ulich Supercomputing Centre,\\
Forschungszentrum J\"ulich, 52425 J\"ulich, Germany}
\affiliation{AIDAS, 52425 J\"ulich, Germany}
\author{Fengping Jin}
\affiliation{Institute for Advanced Simulation, J\"ulich Supercomputing Centre,\\
Forschungszentrum J\"ulich, 52425 J\"ulich, Germany}
\author{Kristel Michielsen}
\affiliation{Institute for Advanced Simulation, J\"ulich Supercomputing Centre,\\
Forschungszentrum J\"ulich, 52425 J\"ulich, Germany}
\affiliation{AIDAS, 52425 J\"ulich, Germany}
\affiliation{RWTH Aachen University, 52056 Aachen, Germany}
\author{Hans De Raedt}
\affiliation{Institute for Advanced Simulation, J\"ulich Supercomputing Centre,\\
Forschungszentrum J\"ulich, 52425 J\"ulich, Germany}
\affiliation{Zernike Institute for Advanced Materials, University of Groningen, Nijenborgh 4, 9747 AG Groningen, The Netherlands}

\date{\today}

\begin{abstract}
  We study large-scale applications using a GPU-accelerated version of the massively parallel J\"ulich universal quantum computer simulator (JUQCS--G).
  First, we benchmark JUWELS Booster, a GPU cluster with 3744 NVIDIA A100 Tensor Core GPUs. Then, we use JUQCS--G to study the relation between quantum annealing (QA) and the quantum approximate optimization algorithm (QAOA). 
  We find that a very coarsely discretized version of QA, termed approximate quantum annealing (AQA), performs surprisingly well in comparison to the QAOA. It can either be used to initialize the QAOA, or to avoid the costly optimization procedure altogether.
  Furthermore, we study the scaling of the success probability when using AQA for problems with 30 to 40 qubits. We find that the case with the largest discretization error scales most favorably, surpassing the best result obtained from the QAOA.
\end{abstract}


\maketitle

\section{Introduction}

The simulation of universal quantum computers requires a large number of matrix-vector updates, most of which are 2-component and 4-component tensor operations. As such, the task of simulating quantum computers is an ideal candidate to profit from recent developments in the GPU industry. We use a GPU-accelerated version of our in-house software JUQCS~\cite{deraedt18, Willsch2020BenchmarkingWithJUQCS}, termed JUQCS--G, to benchmark JUWELS Booster, a cluster of 3744 NVIDIA A100 Tensor Core GPUs, integrated in the modular supercomputer JUWELS~\cite{JUWELS}. A dockerized version of JUQCS is available online \cite{JUQCSDocker}.

JUWELS Booster is part of the JUWELS cluster-booster architecture at the Jülich Supercomputer Centre (JSC) in which a cluster of multi-core nodes is connected via a high-speed network to a cluster of GPUs, the booster, which forms the basis of the modular supercomputer at JSC. The modular supercomputer architecture generalizes the cluster-booster concept by potentially interconnecting a variety of modules with, among others, different acceleration technologies, AI-adapted nodes and storage devices. The modular supercomputer concept allows for a seamless integration of quantum computing architectures and future neuromorphic systems to realize the vision of a holistic future hybrid supercomputer \cite{Suarez2020DevelopingExascaleJSC}. Such a system enables hybrid simulations involving quantum and/or neuromorphic devices that open up new possibilities for demanding computing tasks in science and industry. This will eventually allow for hybrid computing paradigms in a production environment.

JUQCS is a massively parallel simulator \cite{deraedt07, deraedt18, Willsch2020BenchmarkingWithJUQCS, JUQCSDocker} that has also been used for Google's quantum supremacy demonstration \cite{Google2019QuantumSupremacy}.
Using JUQCS--G, we study the quantum approximate optimization algorithm (QAOA) \cite{farhi14, Farhi2016QuantumSupremacyThroughQAOA}, a popular variational algorithm for near-term gate-based quantum computers, also known as noisy intermediate-scale quantum (NISQ) devices \cite{Preskill2018NISQ}. The prospect of producing useful results for NISQ devices has stimulated considerable interest in the scientific community \cite{wang18,OtterbachRigetti2017unsupervised19qubits, Qiang2018QAOA, willsch20_qaoa, Vikstal2019QAOATailAssignment, Bengtsson2020QAOAExactCoverProblem,lacroix20,pagano19, zhou20, Akshay2020QAOAReachabilityDeficit, Harrigan2021QAOASycamore, FernangezPendas2021QAOAClassicalMinimizers, Medvidovic2021QAOASimulation54qubits}.

The QAOA simulations, which were performed on the JUWELS Booster, used the CPUs to carry out the classical (optimization) part of the QAOA and the GPUs to carry out the quantum part formulated in terms of a quantum circuit. On the modular supercomputer architecture with a quantum module, the optimization could be performed on the CPUs of the JUWELS cluster or booster and the operations in the quantum circuit on the QPUs (quantum processing unit), enabling efficient quantum-classical hybrid computations.

The QAOA can be related to a discretized version of quantum annealing (QA)~\cite{willsch20_qaoa,zhou20,streif20,sack2021QAInitializationQAOA}. QA is another popular paradigm of quantum computation \cite{Apolloni89,finnila94,kadowaki98,Brooke99, harris10_eightqubit, Johnson2011DWave, bunyk14, Job2018TestDriving1000Qubits, Hauke2020PerspectivesQuantumAnnealing, Nath2021QMLForRealWorlApplications} that is studied alongside the gate-based model of quantum computation \cite{NIEL10}. Special devices built to perform QA are the D-Wave quantum annealers. The largest existing quantum annealer is the D-Wave Advantage, which has 5000+ physical qubits \cite{dwave2020Advantage}
and has been used for quantum support vector machines \cite{Bhatia2021PerformanceAnalysisQSVM,Phillipson2021QSVMAdvantageBenchmark} (see also~\cite{Willsch2020QSVM}), in studies of stock markets~\cite{cohen2020picking}, for computer vision~\cite{birdal2021QuantumSync}, and for lattice gauge theory~\cite{rahman2021su2}. 
It has recently been benchmarked with 3D spin glass problems~\cite{king2020performance}, garden optimization problems \cite{calaza2021gardenoptimization} and exact cover problems \cite{Willsch2021BenchmarkAdvantage}.
In the present work, the same exact cover problems as in Ref.~\onlinecite{Willsch2021BenchmarkAdvantage}, derived from simplified optimization problems encountered in airplane scheduling, are used to analyze the large-scale simulation results produced by different physical models designed to solve such problems.

In this paper, we scrutinize the overlapping region between QA and the QAOA. We start from a coarse, second-order time-discretization of QA that we call \emph{approximate quantum annealing} (AQA).
We increase the time step that controls the discretization error (sometimes referred to as the \emph{Trotter error} \cite{Heyl2019TrotterError, Sieberer2019TrotterError}, although the formalism goes well beyond Trotter's investigation \cite{Trotter1959Formula}, see \cite{suzuki1976GeneralizedTrotterFormula, DeRaedt83GeneralizedTrotter, Suzuki1985ProductFormulaError}). Furthermore, we use JUQCS--G to study the scaling of the success probability when using AQA for exact cover problems with 30 to 40 qubits. Surprisingly, we find that, while the cases with smaller discretization error provide useful initializations for the QAOA, the cases with largest discretization error scale much better when increasing the number of qubits.

Ideas that are similar to AQA have been investigated before~\cite{zhou20,streif2019comparisonQAOAQASA,streif20,sack2021QAInitializationQAOA}. In particular, in
\cite{sack2021QAInitializationQAOA} a first-order discretized version of QA, referred to as \emph{Trotterized quantum annealing}, was used as initialization for the QAOA. The authors studied the performance for $p\le 10$ QAOA steps and relatively small systems with $N\le 12$ qubits.
Here, we study a \emph{second-order} discretization of QA. We study not only the QAOA initialization but also the \emph{dynamics} of AQA. Furthermore, we consider much larger systems with up to $N=40$ qubits and up to $n=100$ steps (corresponding to $p=101$). 

While it is almost trivial to simulate short QAOA gate circuits for less than 26 qubits on a modern PC, simulating the fairly lengthy circuits (5000+ gates) for the 40 qubits exact cover problems requires substantial supercomputer resources (and more than 16TB of random access memory). The GPU-enabled software that we have developed in house enables us to perform such simulations in a reasonable time span. Having data for 30--40 qubits allows us to assess the potential, e.g.~the scaling behavior, of the QAOA and AQA in a regime that was previously inaccessible (in practice). 

This paper is structured as follows. 
In Section \ref{sec:simulation}, we describe the GPU-accelerated universal quantum computer simulator JUQCS--G and show benchmarks of JUWELS Booster.
In Section \ref{sec:applications}, we present applications to QA, AQA, and the QAOA. We summarize our findings in Section~\ref{sec:conclusion}.

\section{JUQCS--G}
\label{sec:simulation}
In this section, we outline the central task performed by universal quantum computer simulators such as JUQCS in general, and its GPU-accelerated version JUQCS--G in particular. After this, we present benchmark results for JUWELS Booster.

\subsection{Simulating quantum computers on GPUs}
\label{sec:juqcsg}

\begin{figure*}
  \centering
  \includegraphics[width=\textwidth]{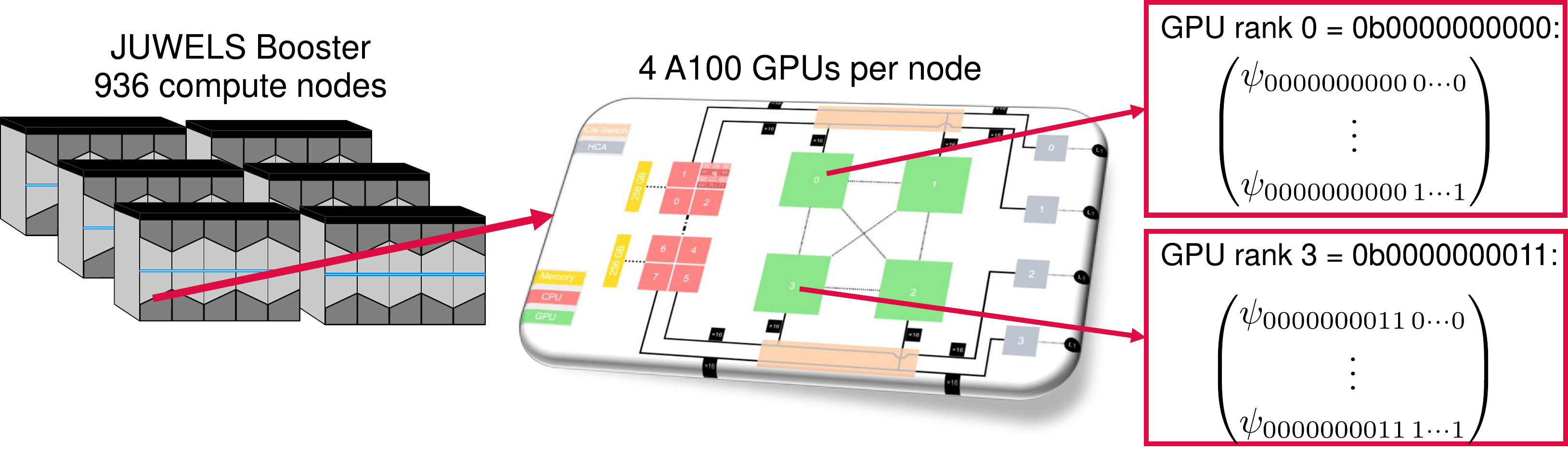}
  \caption{\textbf{Distribution of the complex amplitudes of the state vector $\ket\psi$ on the GPUs across the compute nodes.} Each GPU is handled by one MPI process. For each GPU, the leftmost qubit indices of the coefficients (the \emph{global} qubits, separated from the \emph{local} qubits by a space) represent the MPI rank that uniquely identifies the GPU in the supercomputer. This is indicated for the GPUs belonging to
  MPI rank 0 and 3 for a case with 10 global qubits. On each GPU, the complex amplitudes
  for each index of the remaining local qubits are stored.  During non-local quantum gate operations, typically half of all
  complex amplitudes need to be transferred once between $N_{\mathrm{GPU}}/2$ pairs of GPUs (often across different compute nodes). For these transfers, the MPI communication scheme of JUQCS--G follows the original one described in \cite{deraedt07},
  with the only qualitative change being that a CUDA-aware MPI implementation is used to transfer memory between the GPUs.}
  \label{fig:mpi}
\end{figure*}

The basic unit of computation for a gate-based quantum computer is a single qubit, described by two complex numbers $\ket\psi = (\psi_0, \psi_1)$ that are normalized so that $\braket{\psi|\psi}=|\psi_0|^2+|\psi_1|^2=1$. By definition, an $N$-qubit system  is described by $2^N$ complex numbers 
\begin{eqnarray}
\ket\psi&=&\psi_{0\ldots00}\ket{0\ldots00}+
\psi_{0\ldots01}\ket{0\ldots01}+\ldots
\nonumber \\
&&+\psi_{1\ldots11}\ket{1\ldots11}
\;,
\end{eqnarray}
where $\ket{0\ldots00},\ldots,\ket{1\ldots11}$ are the computational basis states~\cite{NIEL10} and the coefficients $\psi_{0\ldots00},\ldots,\psi_{1\ldots11}$ are
normalized such that $\braket{\psi|\psi}=1$.
For clarity, we explicitly write the $2^N$ complex coefficients in the state 
$\ket\psi$ as a rank-$N$ tensor $\psi_{q_{N-1}\cdots q_1 q_0}$ with indices $q_j\in\{0,1\}$.
In other words, an $N$-qubit system is described by a complex-valued, rank-$N$ tensor, a tensor product of $N$ two-dimensional vectors.

For large-scale universal quantum computer simulations, the main difficulty lies in the management of all $2^N$ complex numbers. For instance, for $N=42$ using double precision floating-point numbers, the tensor $\psi_{q_{N-1}\cdots q_1 q_0}$ occupies  $16\times2^{42}\,\mathrm B = 64\,\mathrm{TiB}$ of distributed memory.

JUQCS--G distributes the complex numbers over the memory of the GPUs as indicated in Fig.~\ref{fig:mpi}. Each GPU stores $2^M$ coefficients of $\ket\psi$ in its local memory, i.e., each GPU stores the coefficients $(\psi_{q_{N-1}\cdots q_M0\cdots0},\,\ldots,\,\psi_{q_{N-1}\cdots q_M1\cdots1})$. For this reason, we call the rightmost $M$ qubits $q_{M-1}\cdots q_{0}$ \emph{local} qubits. As a consequence, the total number of required GPUs is given by $N_{\mathrm{GPU}}=2^{N-M}$. 

Since the complex numbers are distributed over multiple GPUs on different compute nodes, data has to be transferred over the network. This is necessary, for instance, if a particular part of the data on one node is required for the computation on another node. To exchange data between the GPUs on different compute nodes, the \emph{Message Passing Interface} (MPI) is used. MPI provides a standard for distributed memory computation and takes care of the communication, i.e., the sending and receiving of data between different compute nodes. For details, we refer the reader to the literature \cite{mpi40}.

JUQCS--G uses CUDA-aware MPI to manage the distributed memory. Each GPU is controlled by one MPI process, whose rank $r\in\{0,\ldots,N_{\mathrm{GPU}}-1\}$ is initially given by the leftmost $N-M$ indices of $\ket\psi$ in binary notation. Thus, the GPU with rank $\mathrm{bin}(r)=q_{N-1}\cdots q_M$ holds the coefficients $(\psi_{\mathrm{bin}(r)0\cdots0},\,\ldots,\,\psi_{\mathrm{bin}(r)1\cdots1})$. For this reason, we call the leftmost $N-M$ qubits \emph{global} qubits.

A quantum gate is a unitary operation that transforms some of the coefficients of $\ket\psi$. The most elementary quantum gate is a single-qubit gate, i.e., a $2\times2$ unitary matrix $U = (u_{qq'})$. It transforms the coefficients of $\ket\psi$ in terms of 2-component updates. For instance, a single-qubit gate on qubit $j$ transforms the tensor $\ket\psi$ according to
\begin{align}
    \label{eq:twocomponentupdates}
    \psi_{q_{N-1}\cdots q_{j+1}qq_{j-1}\cdots q_0} \leftarrow \sum_{q'=0}^1 u_{qq'} \psi_{q_{N-1}\cdots q_{j+1}q'q_{j-1}\cdots q_0},
\end{align}
for $q=0,1$. Similarly, a two-qubit gate is a $4\times 4$ unitary matrix that operates on two indices of $\ket\psi$, and a three-qubit gate operates on three indices, etc. 
A suitable set of one- and two-qubit gates suffices to construct a universal quantum computer (simulator)~\cite{Deutsch95universality, divincenzo1995twoqubitgates}.
The set of quantum gates implemented by JUQCS--G is documented in \cite{deraedt18}.

We do not use sparse matrix techniques but exploit the special structure of single-, two- and three-qubit operations. We never store or operate on large dense matrices.
To perform 2-component updates as expressed in Eq.~(\ref{eq:twocomponentupdates}), we loop over all pairs of elements of $\ket{\psi}$ and multiply each pair of elements with the same $2\times2$ matrix (which depends on the particular gate). The grouping in pairs depends on the qubit that is being operated on. For the two-qubit operations, we loop over quadruples of elements of $\ket{\psi}$ and multiply each quadruple of elements with the same $4\times4$ matrix (which depends on the particular gate). The time it takes to perform all these arithmetic operations is counted as ``compute time''.
As the size of the quantum computer increases, we need more and more compute nodes to store $\ket{\psi}$, and although the MPI communication is very efficient by itself, it takes an increasingly larger part of the elapsed time (but still scales approximately linearly, not exponentially).

If a quantum gate acts on a global qubit, coefficients of $\ket\psi$ that are stored on different GPUs need to be combined with each other. This requires MPI communication between the GPUs. 
For circuits with many quantum gates involving global qubits, the MPI communication may take a large part of the simulation time (cf.~Fig.~\ref{fig:benchmark_h} below).
For instance, a single-qubit gate on a global qubit requires the transfer of $2^N/2$ complex numbers (i.e., half of all memory) between pairs of GPUs. JUQCS--G minimizes the communication overhead by \emph{relabeling} global and local qubits after such a global quantum gate. Thereby, the complex numbers need to be transferred over the network only \emph{once}, and not back again after the transformation.
Each GPU keeps track of the labeling of global and local qubits in a local permutation array.
Further details of this optimal MPI communication scheme are explained in \cite{deraedt07}.

The keyword in the large-scale simulations performed by JUQCS is \emph{universal}. It means that any quantum circuit for an $N$-qubit system can in principle be simulated, as long as the circuit depth is not unreasonably long (\emph{unreasonably} because then it would also not be executable on a gate-based quantum computer device).
In the literature, this kind of simulation method is sometimes referred to as the \emph{Schr\"odinger simulation method}, because the whole tensor $\psi_{q_{N-1}\cdots q_1 q_0}$ (i.e., the whole \emph{wave function} $\ket\psi$) is propagated through the quantum circuit. The simulation time grows linearly in the total number of gates.

In contrast to the Schr\"odinger simulation method, there is also the so-called \emph{Feynman simulation method} \cite{PEDN17, BOIX17, CHEN18, MARKOV18, VILLA19, Villalonga2019qFlex}. Here, tensor networks are used to obtain only one (or a few) amplitude(s) of the final quantum state. One then sums over each path through the quantum circuit that would contribute to this amplitude. In principle, much larger qubit systems can then be simulated (e.g., a 128-qubit circuit was simulated in \cite{Willsch2020BenchmarkingWithJUQCS}). Of course, the kinds of circuits that can be simulated by such an approach are very restricted and not universal. The simulation time grows exponentially in the circuit depth and depends strongly on the number of Schmidt coefficients of multi-qubit gates (see the supplementary material of \cite{Google2019QuantumSupremacy}). However, truncating Schmidt coefficients opens the possibility to simulate circuits with smaller fidelity. An overview of the limits of such simulations is given in \cite{Zhou2020WhatLimitsSimulationQC}.

A combination of both Schr\"odinger and Feynman approaches can be used to simulate larger circuits of the quantum supremacy experiment \cite{Google2019QuantumSupremacy}, and has recently been used on a cluster of GPUs to spoof the quantum supremacy test
\cite{pan2021simulating}. 

\subsection{Benchmarks and scalings}
\label{sec:benchmark}

The large amount of MPI communication required for simulating universal quantum computations makes simulating quantum computers an ideal candidate to benchmark large supercomputers. Combined with the many  tensor operations required (cf.~Section \ref{sec:juqcsg}), JUQCS--G is a very versatile application to benchmark Tensor Core GPUs. In this section, we report benchmark results for JUQCS--G running on JUWELS Booster, a cluster with $3744$ NVIDIA A100 Tensor Core GPUs distributed over 936 compute nodes (see Fig.~\ref{fig:mpi}).

Each A100 GPU has a local memory of $40\,\mathrm{GiB}$, so the maximum number of local qubits is $31$. For quantum circuits with $N\ge32$ qubits, MPI communication between the GPUs is necessary. For the present benchmark study, we simulate quantum circuits for $32$--$42$ qubits on $2$--$2048$ GPUs.

\begin{figure*}
  \centering
  \includegraphics[width=\textwidth]{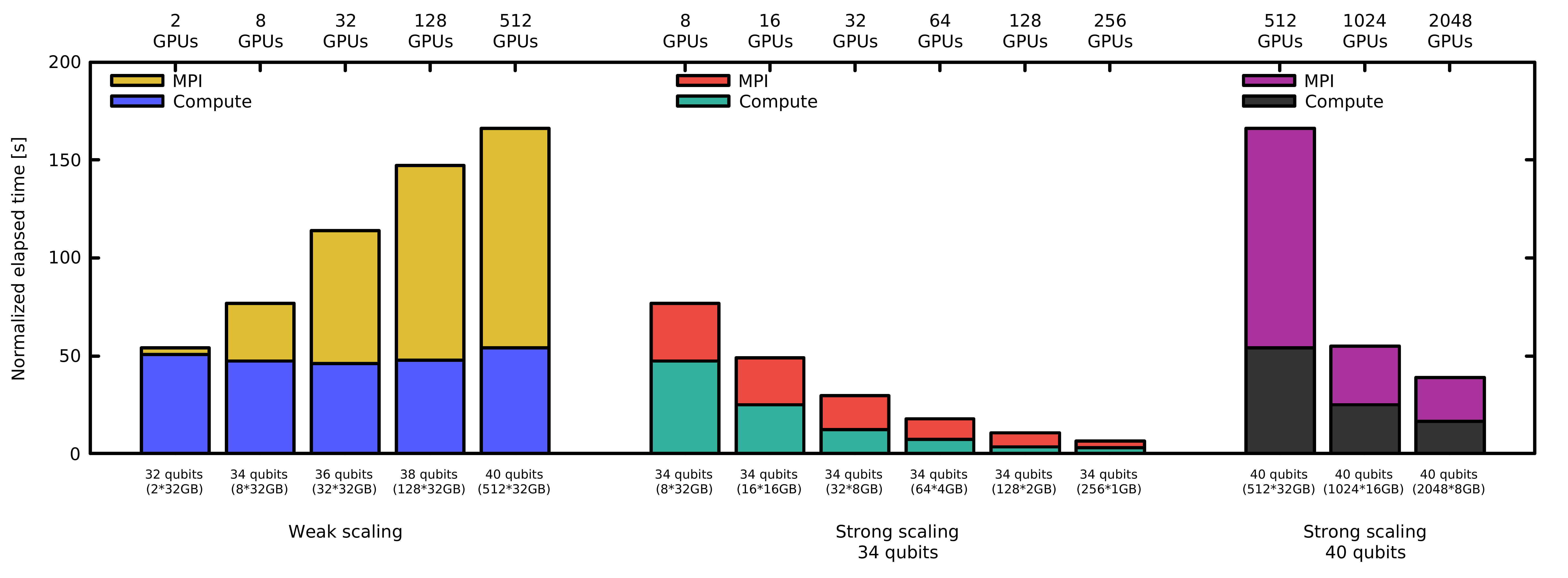}
  \caption{\textbf{Weak and strong scaling results for QAOA on JUWELS Booster using 4 NVIDIA A100 GPUs per node.} Shown is the normalized elapsed time given by Eq.~(\ref{eq:normalizedtime}) as a function of the number of GPUs. The problem size given by the number of qubits and the memory per GPU are indicated on the bottom axis. ``Compute'' refers to the elapsed time
  for executing the quantum circuit only. ``MPI'' refers to the elapsed time for communication plus the elapsed time to prepare
  and postprocess MPI buffers. There is no overlap between computation and communication.}
  \label{fig:benchmark_ec}
\end{figure*}

In Fig.~\ref{fig:benchmark_ec}, we show simulation results for QAOA circuits for $32$--$40$ qubit exact cover problems (the details of which are described in the following section). We see that the computation time (i.e., the run time excluding the time required for the MPI communication) stays approximately constant with increasing system size, indicating ideal weak scaling. The MPI communication time increases roughly linearly. Most importantly, none of these simulation times grow \emph{exponentially} in the number of qubits. In this sense, JUQCS--G beats the exponential growth associated with quantum circuit simulations.

To compare timing data of different runs of problems belonging to the same class, it is expedient to express this data in a way that takes into account that the number of gates depends on the problem size $N$.
In the present case, we take the number of gates $n_{\mathrm{gates}}(32)$ for the smallest corresponding problem instance as reference and define 
\begin{equation}
    \label{eq:normalizedtime}
    \hbox{Normalized elapsed time}=
    \frac{n_{\mathrm{gates}}(32)}{
    n_{\mathrm{gates}}(N)}\, T_{\mathrm{elapsed}}(N)\;.
\end{equation}
Studying the strong scaling results for $34$ qubits, we find ideal strong scaling. As the number of GPUs increases, the normalized elapsed time decreases exponentially. When doubling the number of GPUs used, the normalized elapsed time is (almost perfectly) halved.

Looking closely at the $40$-qubit strong scaling results in Fig.~\ref{fig:benchmark_ec} (rightmost bars), we see that the drop in simulation time from 512 to 1024 GPUs is in fact better than expected. For perfect strong scaling, we would expect the simulation time to decrease by a factor of 2 when doubling the number of GPUs (in practice, this decrease would be expected to be even a little less). Going from 512 to 2048 GPUs, i.e., using 4 times as many GPUs, brings the normalized elapsed time down by almost a factor of 4 as expected. This holds for the computing time as well as for the MPI communication time. However, we observe that the time needed with 1024 GPUs is only a third of the time needed with 512 GPUs, so much better than the theoretical optimum. Note that the unexpected behavior can be attributed to the MPI communication part only. Considering only the computing time, we still observe the expected scaling. 
As this run was performed in October 2020 during the early testing period of JUWELS Booster, we assumed that an explanation for the behavior might be found in an irregularity in the DragonFly+ topology of the communication network. 

\begin{figure*}
  \centering
  \includegraphics[width=\textwidth]{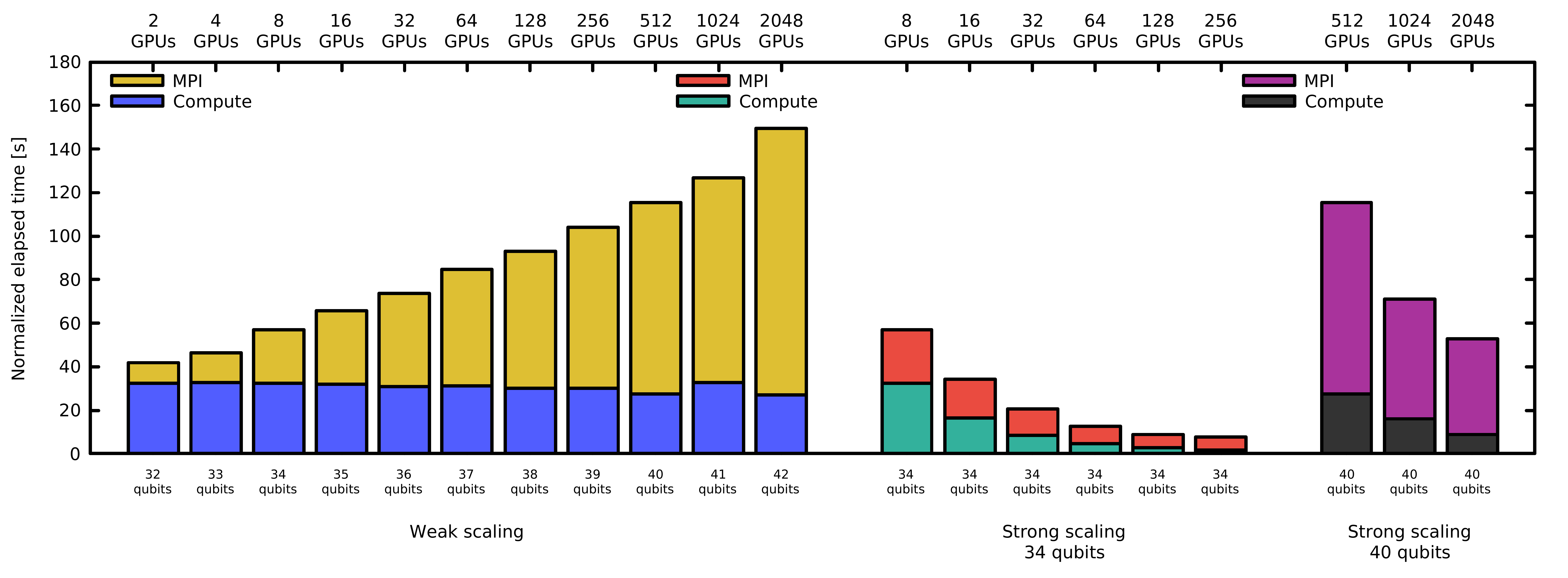}
  \caption{\textbf{The same as Fig.~\ref{fig:benchmark_ec} but for the Hadamard benchmark circuits $(H^{\otimes N})^{11}$.} In this case, the largest runs for 40--42 qubits were repeated several times to estimate the fluctuations due to different node allocations; they were on the order of 1 second and thus negligible (data not shown).
   ``Compute'' refers to the elapsed time
  for executing the quantum circuit only. ``MPI'' refers to the elapsed time for communication plus the elapsed time to prepare
  and postprocess MPI buffers. There is no overlap between computation and communication.}
  \label{fig:benchmark_h}
\end{figure*}

Therefore, we repeated the large-scale benchmark in February 2021 after JUWELS Booster went into production. This time, we used quantum circuits consisting only of Hadamard gates on each qubit, repeated 11 times in a row, $(H^{\otimes N})^{11}$. Such circuits have been found to be well-suited for both benchmarking gate-based quantum computers \cite{Michielsen2017BenchmarkingQC} and universal quantum computer simulators \cite{deraedt18}. They create uniform superpositions over all $N$ qubits and require exchanging $2^N/2$ complex numbers over the whole GPU network for each global single-qubit $H$ gate. Since the total number of gates as a function of $N$ is not constant, we need to make the benchmark results for different $N$ relatable by normalizing the run times w.r.t.~the 32-qubit version. For instance, as the 32-qubit circuit has 352 $H$ gates and the 42-qubit circuit has 462 $H$ gates, the corresponding normalization factor is given by $462/352\approx1.31$. The 11-fold repetition of the Hadamard gates makes potential GPU/CUDA/MPI initialization times negligible.

The results of this second benchmark are shown in Fig.~\ref{fig:benchmark_h}. We see that in this case, the computation times show nearly ideal scaling, i.e., the elapsed time for increasing qubit number and number of used GPUs stays approximately constant (ideal weak scaling) and for constant qubit numbers, doubling the number of used GPUs halves the computation time (ideal strong scaling) in the 34-qubit case as well as in the 40-qubit case. Also for the MPI communication time, the results follow the theoretical expectation.

To compare the speedup over the CPU-based version of JUQCS, JUQCS--E \cite{deraedt18}, we also report results for the normalized run times for the largest circuits in Table~\ref{tab:benchmarkcoresonly} using only CPUs. In this mode of operation, JUWELS Booster can also run 43-qubit circuits.

For the 42-qubit case, we see that the normalized run time on 2048 CPUs, $t_{\mathrm{total}}=2632.4\,\mathrm s$, is a factor of 18 larger than the GPU-accelerated version with $t_{\mathrm{total}}=149.4\,\mathrm s$ (also shown in Fig.~\ref{fig:benchmark_h}). Furthermore, after subtracting the MPI communication time $t_{\mathrm{MPI}}$, the speedup due to the GPU acceleration for the computation-only part is 49. 
This is a very significant improvement in terms of the computational resources required for the simulations.
Clearly, large-scale quantum circuit simulations can tremendously benefit from recent GPU developments.

\begin{table}[tb]
  \caption{\textbf{Comparison of the GPU-based simulator JUQCS--G (first row) and the CPU-based simulator JUQCS--E \cite{deraedt18} (second to last row) for the largest systems using the Hadamard benchmark circuits $(H^{\otimes N})^{11}$.}
  The time $t_{\mathrm{total}}$ is the run time spent for the
  total simulation, normalized by the number of gates with respect to the 32-qubit case (see \equref{eq:normalizedtime}). The time $t_{\mathrm{MPI}}$ is the elapsed time for communication plus the elapsed time to prepare and postprocess MPI buffers.
  JUQCS--E uses all cores of the CPUs on each node.}
   \begin{center}
    \begin{ruledtabular}
      \begin{tabular}{ccccccc}
       qubits& nodes& processes & hardware & normal. &   $t_{\mathrm{total}}\,[s]$ & $t_{\mathrm{MPI}}\,[s]$\\
        \colrule
        42   &  256 & 1024  & GPU & 1.31 & 149.4         &    122.1 \\
        42   &  256 & 2048  & CPU & 1.31 &  2632.4     &    1297.7  \\
        42   &  512 & 4096  & CPU & 1.31 &  1500.4     &    763.4  \\
        43   &  512 & 4096  & CPU & 1.34 &  2714.4     &    1343.3  \\
      \end{tabular}
      \end{ruledtabular}
    \label{tab:benchmarkcoresonly}
  \end{center}
\end{table}

\section{Applications}
\label{sec:applications}

In this section, we use JUQCS--G to study the quantum computer applications QA, AQA, and the QAOA. 
The QAOA work presented is, in spirit, similar to the work reported in
Ref.~\onlinecite{willsch20_qaoa}.
However, the largest optimization problems (16 variable MaxCut and 18 variable 2-SAT) studied in
Ref.~\onlinecite{willsch20_qaoa}
are much smaller (recall the exponential dependence on the number of variables) than the 40-variable exact cover problems studied in the present manuscript.

We first outline the mathematical background and its implementations, and then present the simulation results.

\subsection{Background}
\label{sec:background}

In this section, we discuss the methods that we used in our studies. First, we 
briefly review the most important aspects of QA and the QAOA in Sections~\ref{sec:QA} and~\ref{sec:QAOA}, respectively. The definition of the exact cover problem, which is the class of problems that we study in this paper, is given in Section~\ref{sec:exact_cover}.

\subsubsection{Quantum Annealing}\label{sec:QA}
QA was initially intended as an algorithm for conventional computers~\cite{Apolloni89,finnila94,kadowaki98}. Over time, it has evolved into the idea of a quantum computing device that works fundamentally different from the gate-based quantum computer.

\begin{figure}
  \centering
  \includegraphics[width=\columnwidth]{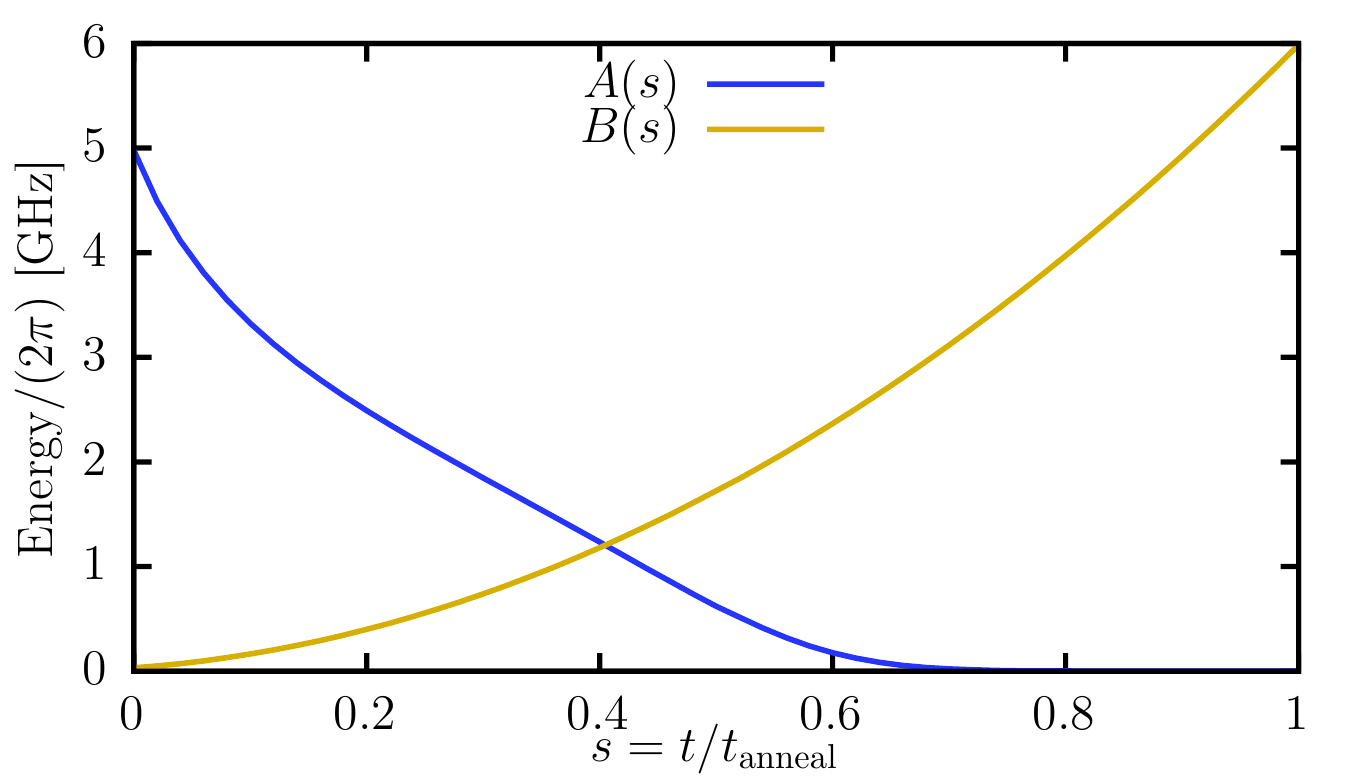}
  \caption{\textbf{Annealing schedule of the DW\_2000Q\_6 quantum annealer, taken from \cite{DWaveAnnealingSchedule}.} The annealing functions $A(s)$ (blue line) and $B(s)$ (yellow line) describe the evolution of the QA Hamiltonian given in \equref{eq:Hofs}.}
  \label{fig:annealing_schedule}
\end{figure}

The concept of QA is based on the adiabatic theorem~\cite{farhi00,childs01}. For this reason, a QA device is also called \emph{adiabatic quantum computer}.
Although slightly different concepts are sometimes associated with QA and adiabatic quantum computation, the basic working principle is the same:
The quantum system (consisting of qubits) is prepared in the ground state of an initial Hamiltonian such as
\begin{align}
   \label{eq:HI}
   H_I = -\sum_{i=0}^{N-1} \sigma_i^x,
\end{align}
whose ground state is given by $\ket{\psi_\mathrm{init}}=\ket{{+}}^{\otimes N}$ where $\ket+=(\ket 0+\ket 1)/\sqrt 2$ is the uniform superposition of $\ket0$ and $\ket 1$.
During the time evolution, the Hamiltonian changes according to
\begin{align}
   \label{eq:Hofs}
   H(s) = A(s)H_I + B(s)H_C,\quad s=t/t_\mathrm{anneal},
\end{align}
where $t_\mathrm{anneal}$ is the time used for the annealing process, and the two annealing functions $A(s)$ and $B(s)$ fulfill $A(0)\gg B(0)$ and $A(1)\ll B(1)$. An example annealing schedule is shown in Fig.~\ref{fig:annealing_schedule}, which is also used to initialize the variational QAOA parameters (see below).

The final Hamiltonian in Eq.~(\ref{eq:Hofs}), $H_C$, represents an optimization problem that is to be solved. This means that the ground state of $H_C$ encodes the solution of a certain optimization problem. Here, we choose $H_C$ to be the Ising Hamiltonian
\begin{align}
  \label{eq:HC}
   H_C = \sum_{i=0}^{N-1} h_i\sigma_i^z + \sum_{i<j}J_{ij} \sigma_i^z\sigma_j^z.
\end{align}

The idea is that if the annealing process described by Eq.~(\ref{eq:Hofs}) is carried out at zero temperature and sufficiently slowly so that the adiabatic theorem holds, then the quantum system stays in its instantaneous ground state. Thus, at the end of the annealing process, the quantum system ends up in the ground state of the Hamiltonian $H_C$. Measuring the qubits would then yield the answer to the initial optimization problem.

In practice, on a quantum annealer not only the annealing time $t_\mathrm{anneal}$ determines the probability of success (i.e., the probability that the system ends in its ground state and not in an excited state), but also an environment at finite temperature, control errors and precision limits have an influence on
it~\cite{harris10_eightqubit,dickson13,bian14,amin15,Mishra2018,marshall19,Pearson2019,weinberg20}.

\subsubsection{The Quantum Approximate Optimization Algorithm}\label{sec:QAOA}
The QAOA was introduced by Farhi et al.~\cite{farhi14}. It is a variational method that is suitable for execution on a gate-based quantum computer. The objective is to find the ground state (or a low energy state) of a problem Hamiltonian such as $H_C$ given by Eq.~(\ref{eq:HC}) that represents an optimization problem.
The state that is prepared by the QAOA quantum circuit is given by
\begin{align}
   \ket{{\beta,\gamma}} = \prod_{k=1}^p e^{-i\beta_kH_D}e^{-i\gamma_kH_C}\ket{+}^{\otimes N},
   \label{eq:QAOA_state}
\end{align}
where
$\gamma=(\gamma_1,...,\gamma_p)$ and $\beta=(\beta_1,...,\beta_p)$ are the $2p$ variational parameters that have to be optimized, and $H_D$ is a \emph{mixing Hamiltonian} that is commonly chosen as
$H_D=-H_I$ (cf.~\equref{eq:HI}), i.e.,
\begin{align}
   H_D = \sum_{i=0}^{N-1} \sigma_i^x.
\end{align}
Note that other choices have also been proposed~\cite{hadfield19,zhu20}.

It is worth mentioning that for this choice, the QAOA parameters $\beta_k$ can be reduced to the range $[0,\pi)$. For $\gamma_k$, however, such a periodicity condition depends on the minimum spacing between the eigenvalues of $H_C$. In other words, the range of values for $\gamma_k$ depends on the particular problem instance defined by $h_i$ and $J_{ij}$ (see below).

For a given number of steps $p$, the energy of the optimized variational state ($E_p^*=\min_{\beta,\gamma} \bra{{\beta,\gamma}}H_C\ket{{\beta,\gamma}}$)  is lower than the energy of the optimized variational state with $p-1$ steps \cite{farhi14}. 

However, it has been found that the optimization of the variational parameters can be rather inefficient. Therefore, one often tries to use the observation that the optimal parameters $\beta_k$ and $\gamma_k$ seem to follow certain patterns~\cite{crooks18,brandao18,willsch20_qaoa,zhou20,Vikstal2019QAOATailAssignment,farhi19}. Here, we investigate these patterns in relation to QA and their interpretation as an optimized annealing scheme (see also \cite{zhou20,willsch20_qaoa,sack2021QAInitializationQAOA}).

\subsubsection{Exact Cover}\label{sec:exact_cover}

The exact cover problem is an NP-complete problem \cite{Karp1972KarpsNPCompleteProblems} that has become a popular choice to study optimization using quantum computing systems
\cite{Farhi2001QuantumAdiabaticNPComplete, choi2010adiabaticQuantumAlgorithms, lucas14, Cao2016SetCoverProblemsQA, Sax2020ApproximateApproximationQA, Vikstal2019QAOATailAssignment, Bengtsson2020QAOAExactCoverProblem,lacroix20,Willsch2021BenchmarkAdvantage}. Exact cover problems belong to the class of set covering and partitioning problems that is covered in a vast amount of literature in Operations Research (see e.g.~\cite{Ernst2004StaffSchedulingAndRostering,Tahir2019IntegralColumnGeneration}).

In this paper, we study the instances of exact cover problems used in \cite{Willsch2021BenchmarkAdvantage}. 
In matrix form, they are written as
\begin{align}
  \label{eq:exactcover}
   \min_{x_i=0,1} \sum_{f=0}^{F-1} \left( \sum_{i=0}^{N-1}a_{if}x_i - 1 \right)^2,
\end{align}
where $a\in\{0,1\}^{N\times F}$ is the Boolean problem matrix that defines the exact cover instance, and $x_i$ are the problem variables. Intuitively, the solution $x\in \{0,1\}^N$ of Eq.~(\ref{eq:exactcover}) selects rows of $a$ in such a way that in each column of the selected rows, the entry 1 is \emph{covered exactly once}, and all other entries are 0.

We study exact cover problems with 30 to 40 variables (qubits) and $F=472$ terms. For each problem size, we have four different instances. Problem instances are labeled by their qubit number and an additional label from 0 to 3 in brackets, such as problem $30(0)$.

To find a problem Hamiltonian $H_C$ of the form of \equref{eq:HC}, whose ground state represents the solution to \equref{eq:exactcover}, we replace the problem variables according to 
\begin{align}
    \label{eq:xtosigma}
    x_i\mapsto(1+\sigma_i^z)/2.
\end{align}
Denoting the $-1$ ($+1$) eigenstate of $\sigma_i^z$ as $\ket0$ ($\ket1$), we can represent the problem variable $x_i$ by the qubit state $\ket{x_i}$. Thus, the replacement \equref{eq:xtosigma} yields a diagonal Hamiltonian whose eigenvalues take all possible values of the objective function in \equref{eq:exactcover}. Consequently, the ground state of $H_C$ (i.e., the state with minimum eigenvalue) is the solution to \equref{eq:exactcover}. For this reason, we also define the \emph{success probability} for these problems as the probability to find the system in the ground state (note that for all problem instances that we study in this paper, the ground state is unique \cite{Willsch2021BenchmarkAdvantage}).

After multiplying out the square, the Hamiltonian can be expressed in the form of \equref{eq:HC} plus an additive constant $C$ (see \cite{Willsch2021BenchmarkAdvantage} for the calculation), yielding
\begin{align}
  \label{eq:HChi}
  h_i &= \sum_j \frac 1 2(aa^T)_{ij} - (ab)_i\;, \\
  \label{eq:HCJij}
  J_{ij} &= \frac 1 2(aa^T)_{ij}\;,  \\
  \label{eq:HCC}
  C &= b^T b +  \frac{1}{2}\sum_{i< j} (aa^T)_{ij} + \frac{1}{2}\sum_i  ((aa^T)_{ii} - (2ab)_i)\;,
\end{align}
where $b = (1,\ldots,1)^T$ is an $F$-dimensional vector of ones.

As $a$ and $b$ in Eqs.~(\ref{eq:HChi})--(\ref{eq:HCC}) contain only zeros and ones, we know that $h_i$ and $J_{ij}$ vary at most by half integers. Therefore, the range of values for $\gamma_k$ can be reduced to $[0,2\pi)$ (because $\gamma_k\mapsto\gamma_k+2\pi$ only causes a global phase in the QAOA state in \equref{eq:QAOA_state}). 

For all AQA and QAOA applications (except the grid scan in Fig.~\ref{fig:qaoa_scan} below), however, we rescale the parameters $\{h_i\}$, $\{J_{ij}\}$ and $C$ to a uniform parameter range by dividing them by
\begin{eqnarray}
r=\max\left\{
\max\left[\frac{\max\{h_i\}}{h_{\mathrm{max}}},0\right],
\max\left[\frac{\min\{h_i\}}{h_{\mathrm{min}}},0\right],\right. \nonumber \\
\left. \max\left[\frac{\max\{J_{ij}\}}{J_{\mathrm{max}}},0\right],
\max\left[\frac{\min\{J_{ij}\}}{J_{\mathrm{min}}},0\right]
\right\},\quad
\label{eq:rescale}
\end{eqnarray}
where $h_{\mathrm{max}}=-h_{\mathrm{min}}=2$ and $J_{\mathrm{max}}=-J_{\mathrm{min}}=1$.
Note that the same normalization is also performed when solving such problems on the D-Wave quantum annealer \cite{DWaveSolversParameters, Willsch2021BenchmarkAdvantage}.
This does not change the solutions of the problems. However, it brings the energies of different problem instances on a uniform scale. This in turn improves the optimization of the QAOA parameters, and it also allows the use of the same AQA time step $\tau$ (see below) for different problem instances.

\subsection{Implementations}

In this section, we discuss how quantum physics simulations are used to carry out the QAOA and AQA.

\subsubsection{QAOA}

We initialize the $2p$ variational QAOA parameters $\beta_k$ and $\gamma_k$ in \equref{eq:QAOA_state} according to the second-order Suzuki-Trotter decomposition. This amounts to (see Appendix~\ref{app:initialization})
\begin{align}
   \label{eq:beta_k}
   \beta_k &= -\tau(A(s_{k+1})+A(s_k))/2, & k&=1,...,p-1,\\
   \beta_{p} &= -\tau A(s_{p})/2,\\
   \gamma_k &= \tau B(s_k), & k &= 1,...,p,\label{eq:gamma_k}
\end{align}
where $s_k = (k-1)/(p-1)$ (slightly different from Ref.~\cite{willsch20_qaoa}) and we take $A(s)$ and $B(s)$ from the DW\_2000Q\_6 annealing schedule \cite{DWaveAnnealingSchedule} (see Fig.~\ref{fig:annealing_schedule}).
This procedure is motivated by the relation between the QAOA and QA as discussed in more detail in \cite{willsch20_qaoa} (see also \cite{sack2021QAInitializationQAOA}
where the first-order case is discussed).

Given values for the variational parameters $\beta$ and $\gamma$, JUQCS--G computes the variational state $\ket{{\beta,\gamma}}$ given in Eq.~(\ref{eq:QAOA_state}), where the exponentials $\exp(-i\beta_kH_D)=\prod_i\exp(-i\beta_k\sigma_i^x)$ are computed as a sequence of rotations around the $x$ axis with angle $2\beta_k$, and the exponentials $\exp(-i\gamma_kH_C)$ are computed as a sequence of rotations around the $z$ axis with angle $2\gamma_k$ and controlled-$Z$ gates.

JUQCS--G also computes the probability of the solution state in the variational state $\ket{{\beta,\gamma}}$ and the energy expectation value $E_p(\beta,\gamma)=\bra{{\beta,\gamma}}H_C\ket{{\beta,\gamma}}$. In the optimization phase of the QAOA, this energy is passed to the optimizer (we use several optimizers from the \texttt{scipy} library \cite{scipy}; see below).
The optimizer then proposes new values for the variational parameters which are in turn passed to JUQCS--G.
If this optimization loop does not reach convergence, we use an additional stopping criterion of a maximum of 200 calls to JUQCS--G.

We note that in practice, it is only possible to optimize for the energy $E_p$ and not for the success probability.
Since we use a state-vector simulator and know the ground state, we could in principle also optimize for the probability to observe the ground state.
However, in our benchmark, we consider the realistic situation that we do not know the ground state and thus optimize for the energy.
We compute the probability of the ground state in a given variational state only as a measure of success.

For optimization problems, often the approximation ratio is also considered as a measure for the performance of the QAOA.
In our case, however, only finding the unique ground state is considered as success since none of the excited states encodes a valid solution to the exact cover problem.

\subsubsection{AQA}

\begin{figure*}
  \centering
  \includegraphics[width=\textwidth]{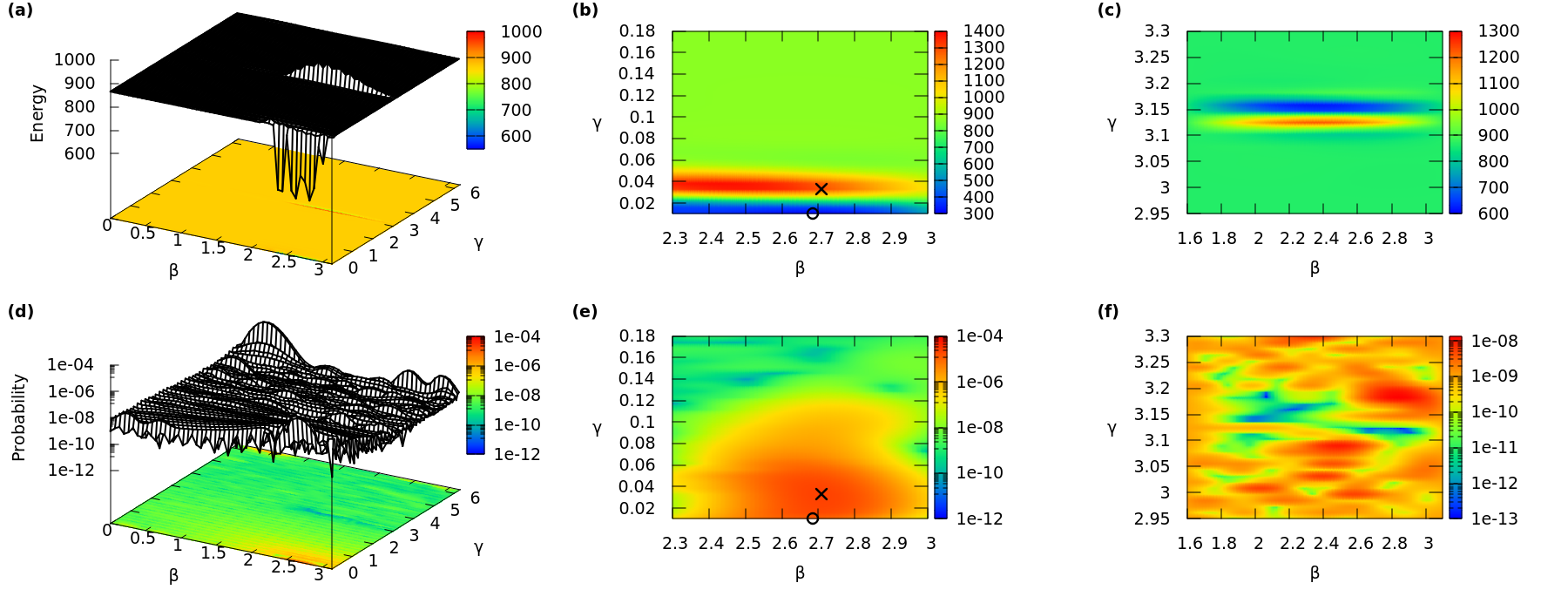}
  \caption{\textbf{(a)--(c) Energy landscape and (d)--(f) success probability landscape for QAOA with $p=1$ for the 30-qubit exact cover problem 30(0).} The left column shows the landscapes obtained by scanning a $64\times64$ grid $\beta\in[0,\pi)$ and $\gamma\in[0,2\pi)$. The middle column shows a zoom into the area around the minimum energy found in the scan. The largest success probability (cross) and the energy minimum (circle) in this area are indicated. Interestingly, these two points are not at the same location. The right column shows a zoom into another area of interest around $\gamma\approx\pi$ and $\beta\lesssim \pi$.}
  \label{fig:qaoa_scan}
\end{figure*}

To introduce the basic idea of AQA, we first review how the time evolution of a QA process is simulated. This allows us to describe in what sense the description becomes ``approximate'', and when the simulation of QA enters the regime of what we call AQA.

In essence, a simulation of QA requires the solution of the time-dependent Schr\"odinger equation (TDSE), $i\partial_t\ket{\psi(t)}=H(t)\ket{\psi(t)}$, with a time-dependent Hamiltonian $H(t)$, such as the QA Hamiltonian given in Eq.~(\ref{eq:Hofs}).
In principle, but also in practice, the time-discretized TDSE can be expressed as a quantum gate circuit which can then be processed by JUQCS--G. 
However, for convenience (and also as a check on the JUQCS data), we often solve the TDSE with the quantum spin dynamics simulator (QSDS) (in house software with the MPI communication scheme taken from JUQCS but without GPU implementation).
QSDS solves the TDSE for the generic spin-1/2 Hamiltonian
\begin{align}
H_\mathrm{QSDS}(t)
&=-\sum_{\mathclap{\alpha=x,y,z}}\;\;
\left(\sum_{i<j} {\widetilde J}^\alpha_{ij}(t) \sigma^\alpha_i \sigma^\alpha_j +
 \sum_{i=0}^{N-1} {\widetilde h}^\alpha_{i}(t) \sigma^\alpha_i\right)
,
\label{AQA0}
\end{align}
where $N$ is the number of spins (qubits).
For the optimization problems at hand, we have
\begin{align}
{\widetilde h}_i^x(t)&=A(t/t_{\mathrm{anneal}}),\\
{\widetilde h}_i^z(t)&=-B(t/t_{\mathrm{anneal}}) h_i,\\
{\widetilde J}^z_{ij}(t)&= -B(t/t_{\mathrm{anneal}}) J_{ij},
\label{AQA3}
\end{align}
where the annealing functions $A$ and $B$ are shown in Fig.~\ref{fig:annealing_schedule}, and $h_i$ and $J_{ij}$ encode the problem instance as before.

QSDS solves the TDSE by time stepping using the second-order Suzuki-Trotter formula~\cite{Suzuki1993GeneralDecompositionTheoryOrderedExponentials,RAED06}
\begin{align}
|\Psi((l+1)\tau)\rangle &=
\left\{\exp\left[\frac{i\tau}{2}\sum_{\alpha=x,z}\;\;
\sum_{i=0}^{N-1} {\widetilde h}^\alpha_{i}(l\tau) \sigma^\alpha_i\right]\right.\nonumber\\
&\times \exp\left[ i\tau \sum_{i<j} {\widetilde J}_{ij}(l\tau) \sigma^z_i \sigma^z_j \right]\nonumber\\
&\left.\times\exp\left[\frac{i\tau}{2}\sum_{\alpha=x,z}\;\; \sum_{i=0}^{N-1} {\widetilde h}^\alpha_{i}(l\tau) \sigma^\alpha_i\right]\right\}
|\Psi(l\tau)\rangle
\;,
\label{AQA4}
\end{align}
for $l=0,\ldots,n$ (such that $t_{\mathrm{anneal}}=(n+1)\tau)$.
Note that the action of each of the matrix exponentials in Eq.~(\ref{AQA4}) on any state vector can be computed exactly.
For the initial state, we take $|\Psi(0)\rangle = |+\rangle^{\otimes N}$.
Apart from collecting all single-spin terms of the Hamiltonian Eq.~(\ref{AQA0})
into the same matrix exponential, the structure of the QAOA (cf.~Eq.~(\ref{eq:QAOA_state})) is the same as that of Eq.~(\ref{AQA4}).

The basic idea of AQA is to solve the TDSE with a time step $\tau$ which is too large to yield an accurate time evolution of a genuine QA process.  Moreover, the number of time steps $n$ is taken to be rather small. Therefore the corresponding ``annealing time'' is rather short in which case the time evolution is unlikely to be adiabatic.

In other words, we do not rely on the adiabatic theorem but hope that with a relatively small number of factors in the product formula
with a relatively large time step, we can nevertheless generate a final
state which is close to the ground state of the problem Hamiltonian.
Clearly, AQA is a heuristic method, partially motivated by
findings~\cite{crooks18,brandao18,willsch20_qaoa,zhou20,Vikstal2019QAOATailAssignment} that optimal values for the variational parameters $\beta_k$ and $\gamma_k$ were often found to follow curves which resemble such an approximate annealing schedule.
For AQA, we use again Eqs.~(\ref{eq:beta_k})--(\ref{eq:gamma_k}) but with the convention: $s_k=k/n$ for $k=0,...,n$.

For each step, AQA and QAOA perform exactly the same number of single-qubit  and two-qubit gates.
Only the single-qubit gates may require exchange of data among MPI processes.
In our AQA simulations, we also compute the spin expectation values during the time evolution.
In terms of computational effort, AQA for a fixed $n$ (i.e.~$n+1$ steps since we start counting at 0) is equivalent to a single evaluation of a QAOA circuit with $p=n+1$.

\subsection{Results}
\label{sec:results}

In this section, we present the simulation results on the QAOA, AQA, and a comparison between them.

\subsubsection{QAOA}
\label{sec:QAOAresults}

\begin{figure}
  \centering
  \includegraphics[width=\columnwidth]{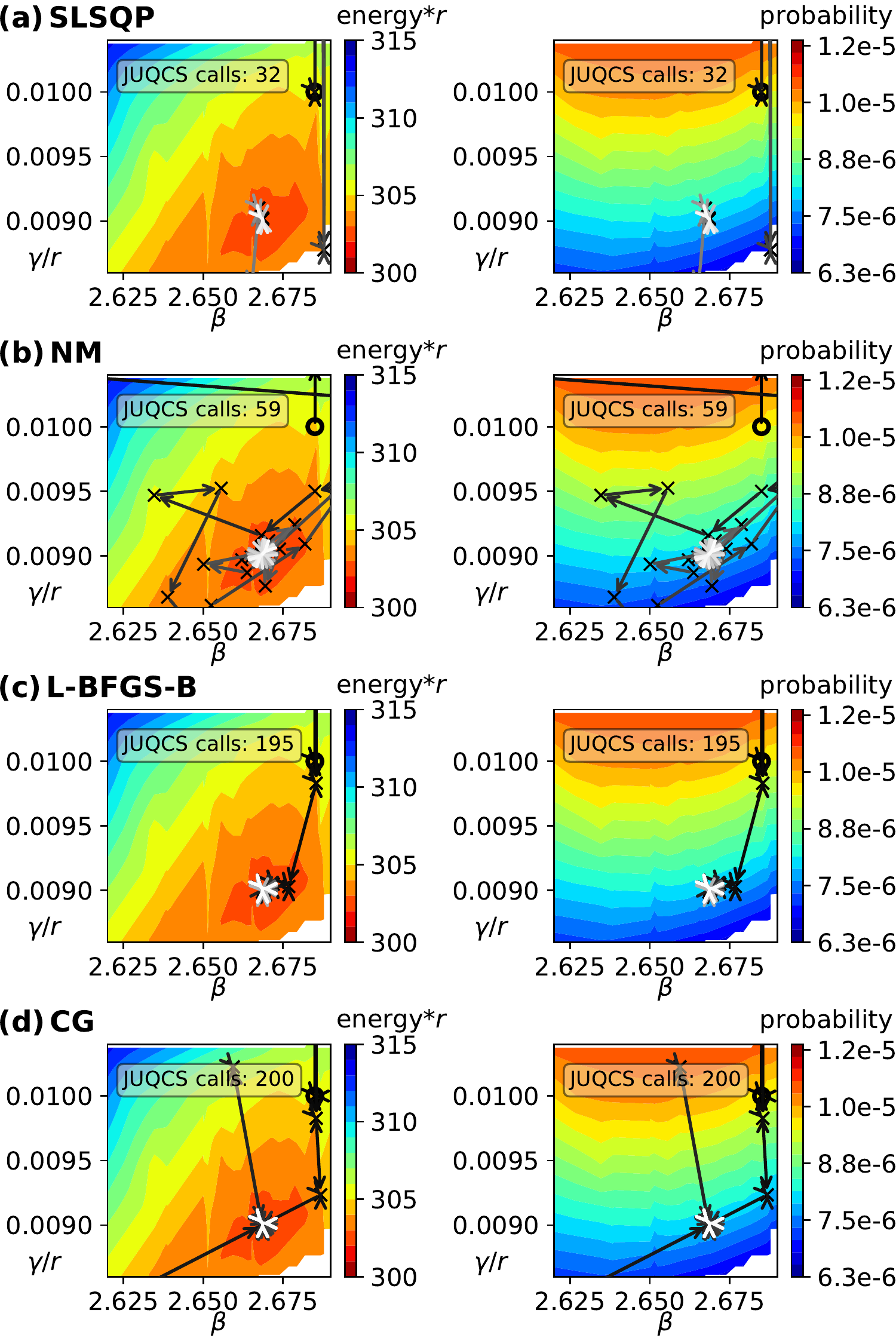}
  \caption{\textbf{Comparison of different classical optimizers used to optimize the variational parameters $\beta$ and $\gamma$ for QAOA with $p=1$ for the 30-qubit exact cover problem 30(0).}
  The optimizers are \textbf{(a)} SLSQP, \textbf{(b)} NM, \textbf{(c)} L-BFGS-B, and \textbf{(d)} CG (see main text).
  The starting point for the optimization is the point with minimal energy found in the initial scan (black circles, corresponding to the black circles in Figs.~\ref{fig:qaoa_scan}(b) and (e)).
  For each optimizer, the left (right) panel shows the energy (success probability) landscape.
  The number of JUQCS--G calls used by each optimizer is indicated in the top left corner of each panel. The parameters for each call are shown with black crosses.
  The order in which the parameters are evaluated by the optimizers is indicated with arrows with colors evolving from black (for the first JUQCS--G call) to white (for the last JUQCS--G call). Note that, although the rescaled version of the problem was used for the optimization (see \equref{eq:rescale}; here $r=36.75$), we plot $\gamma/r$ and $\mathrm{energy}*r$ to make the scale comparable with the grid scan in Fig.~\ref{fig:qaoa_scan}.}
  \label{fig:qaoa_optimization_landscape}
\end{figure}

We start with QAOA for $p=1$ by studying the energy landscape and the success probability for a 30-qubit problem instance (named 30(0)). For this purpose, we perform a scan of the parameters $\beta\in[0,\pi)$ and $\gamma\in[0,2\pi)$ and compute the energy as well as the success probability obtained for the QAOA circuit (note that, as argued in Section~\ref{sec:exact_cover}, these parameter intervals cover the range of different QAOA trial states).
The results are shown in the left column of Fig.~\ref{fig:qaoa_scan} for the energy (top) and success probability (bottom).
The middle column shows zooms with finer grids into regions around the energy minimum and the success probability maximum. The right column shows zooms into another region of interest noticeable in Fig.~\ref{fig:qaoa_scan}(a).

The point with the highest success probability is marked with a cross and the point with the lowest energy is marked with a circle. Although these points are relatively close, the energy is very different. The point with the highest success probability is even close to an energy maximum. However, the point at the energy minimum still has a relatively high success probability. In the right column, which shows the zoom in the vicinity of another local energy minimum, we find that the success probability is quite low (see the scale of the color bars). If, during the optimization process, the minimizer gets stuck in such a local minimum, the probability to observe the ground state will often be very small.

We find that the optimal parameters $\beta^*$ and $\gamma^*$ in this case are large (almost $\pi$) and small (almost 0), respectively, as would be the values for $A(0)$ and $B(0)$ in an annealing scheme.
This is encouraging for our annealing scheme initialization for QAOA with $p>1$.

Figure~\ref{fig:qaoa_optimization_landscape} shows the paths that different optimization algorithms take when starting from the point with minimum energy
found in the scan (the black circle in Fig.~\ref{fig:qaoa_optimization_landscape}(b)).
The optimization algorithms are standard optimizers provided by \texttt{scipy}~\cite{scipy}: sequential least squares programming (SLSQP), the gradient-free Nelder--Mead algorithm~\cite{NelderMead1965} (NM), the L-BFGS-B algorithm~\cite{Zhu1997LBFGSBalgorithm,Morales2011LBFGSBalgorithmImprovement}, and the conjugate gradient algorithm~\cite{numericalrecipes} (CG) (see \cite{FernangezPendas2021QAOAClassicalMinimizers} for a thorough comparison of different optimizers for the QAOA).
Note that for the optimization, we use the rescaled version of the problem according to \equref{eq:rescale} (here $r=36.75$); otherwise small variations in $\gamma$ led to large fluctuations in the energy and no optimizer except NM was able to converge to the energy minimum (data not shown).

Figure~\ref{fig:qaoa_optimization_landscape} shows that with rescaling, all optimizers converge to the energy minimum, although L-BFGS-B and CG require 3--6 times more quantum circuit simulations than SLSQP and NM. Note, however, that the convergence to the energy minimum depends crucially on the good initial point;
other random initial points produced much worse results (data not shown). Furthermore, it is worth noting that none of the optimizers comes across the point
with the largest success probability (the cross in Fig.~\ref{fig:qaoa_scan}(b)); only NM and CG venture once into a region with better success probability (the right panels in Fig.~\ref{fig:qaoa_optimization_landscape}). Obviously, this is not a flaw of the optimizers (which can only optimize for the energy in practice), but rather a deficiency of variational algorithms in general.

We take a number of QAOA steps $p\le13$ and minimize
the cost function $\bra{{\beta,\gamma}}H_C\ket{{\beta,\gamma}}$ w.r.t.~$\beta_k$ and $\gamma_k$,
as one would do for QAOA running on genuine quantum hardware.
The hope is then that by minimizing the cost function, we will
also obtain relatively large values for the success probabilities.
In Table~\ref{tab:qaoa_results}, we present the results for a set of exact cover instances.

The QAOA results for $p=7,13$ are encouraging in the sense that the success probabilities are relatively large, i.e., much larger than $2^{-N}$ which would be the probability to pick the correct solution from a uniform distribution at random.
However, the number of JUQCS--G calls required to obtain such values is also fairly large. The numbers in parentheses indicate the number of JUQCS--G calls corresponding to the highest observed success probability.
Almost all runs have been terminated after 200 JUQCS--G calls (black entries in Table~\ref{tab:qaoa_results}). Runs which were terminated by the minimizer (red entries) have a substantially lower success probability (smaller than $4\%$), suggesting that the minimizer became stuck in a local minimum.

We also performed some QAOA simulations with $p=3$. We observed that for problem instance 30(0), the achieved success probability was smaller by a factor of 10--20 than in the $p=7$ and $p=13$ cases after using a similar number of JUQCS--G calls. For larger problem instances, the minimizer seemed to get stuck in local optima as the obtained success probabilities were smaller than 1\%. We thus concluded that $p=3$ would be too small for larger problem instances and we did not proceed with $p=3$.

\begin{table}[tb]
  \caption{\textbf{QAOA results for exact cover instances, obtained by minimizing the energy expectation value using SLSQP.} 
  QAOA quantum gate circuits were executed using JUQCS--G.
  The success probability is determined by computing the probability of the ground state after each iteration and searching for the iteration number (given in parentheses) for which this probability is largest. 
  The number of JUQCS--G calls was limited to 200.
  Red colored entries: The run was terminated by the minimizer that was probably stuck in a local minimum; black colored entries: The run was terminated when the 200 JUQCS--G calls had been reached.
  For $p=13$, the calculations are too costly to warrant filling all missing entries.}
  \begin{center}
    \begin{ruledtabular}
      \begin{tabular}{ccc}
        qubits     & \multicolumn{2}{c}{success probability (JUQCS--G calls)} \\
        (instance)  & $p=7$& $p=13$ \\
        \hline\noalign{\vskip 4pt}
        30(0) & 0.3398 (165)                 & 0.6214 (187) \\
        30(3) & 0.3708 (196)                 & - \\
        32(0) & 0.2841 (195)                 & - \\
        32(3) & 0.2745 (192)                 & 0.4741 (193) \\
        34(0) & 0.1924 (190)                 & - \\
        34(3) & 0.2251 (196)                 & 0.5075 (187) \\
        36(0) & 0.1081 (191)                 & - \\
        36(3) & 0.1545 (175)                 &  {\color{red}0.0387 (94)} \\
        38(0) & 0.0901 (187)                 & - \\
        38(3) & 0.1200 (174)                 &  {\color{red}0.0159 (124)} \\
        40(0) & {\color{red}0.0068 (71) }                 &  {\color{red}0.0088 (123)} \\
        40(3) & {\color{red}0.0061 (38) }                 & - \\
      \end{tabular}
    \end{ruledtabular}
    \label{tab:qaoa_results}
  \end{center}
\end{table}

\subsubsection{AQA}
\label{sec:AQAresults}

\begin{figure}[tb]
  \centering
  \includegraphics[width=0.9\columnwidth]{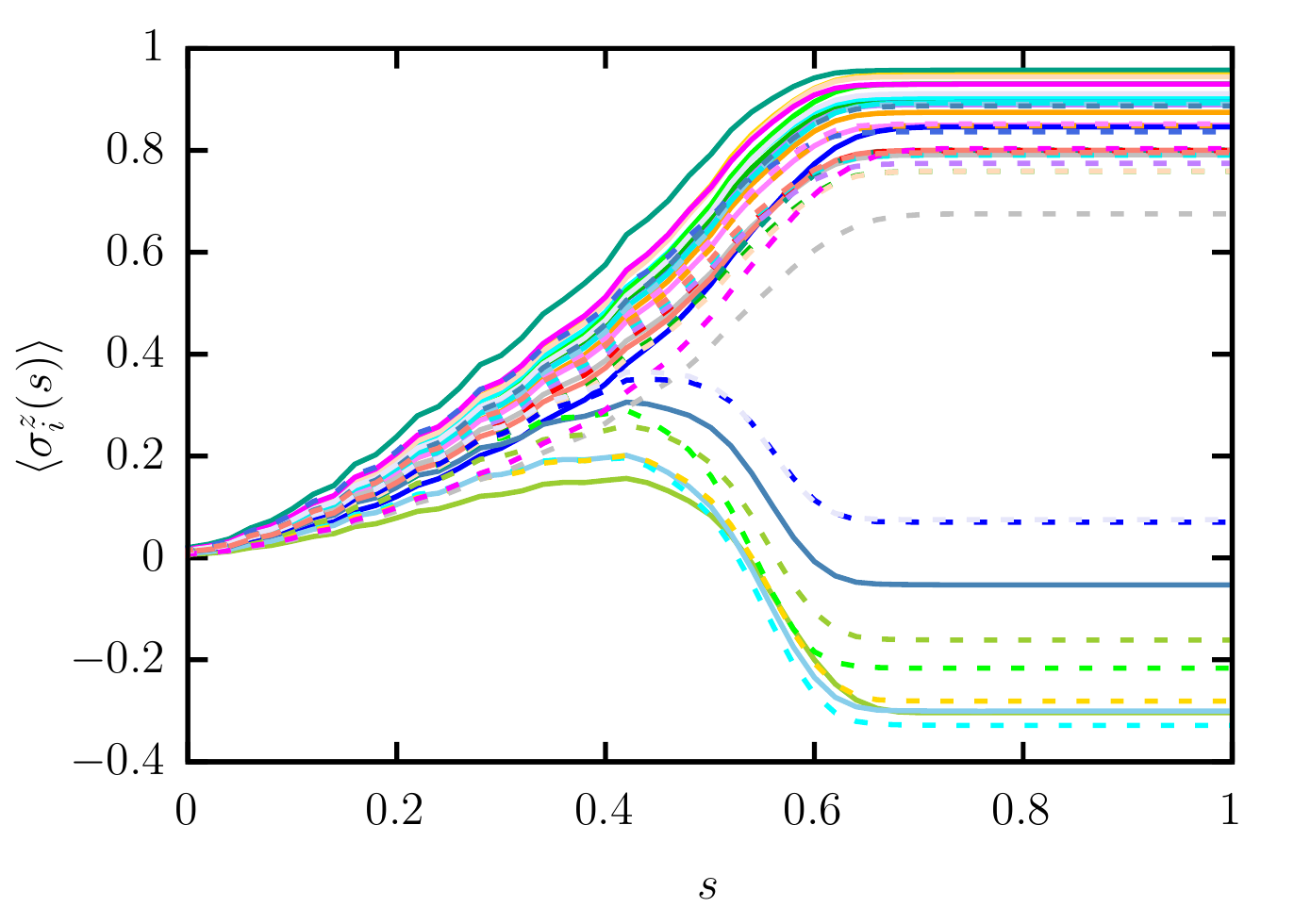}
  \caption{\textbf{AQA results for the 40-variable exact cover instance 40(0).}
  Shown are the spin expectation values $\langle\sigma^z_i(s)\rangle$ during the time evolution generated by \equref{eq:Hofs}, as a function of the normalized annealing time $s=t/t_{\mathrm{anneal}}$.
  Different lines correspond to different qubits $i=0,\ldots,N-1$ for $N=40$.
  The success probability to generate the state corresponding to the solution of this instance is $0.038$. The number of steps is $n=50$ and the time step is $\tau=0.4\unit{ns}$.}
  \label{fig:aqa2}
\end{figure}

A representative AQA result for a 40-variable exact cover problem is shown in
Fig.~\ref{fig:aqa2}.
In this simulation, we chose $n=50$ and the time step $\tau=0.4\,\mathrm{ns}$,
corresponding to a total annealing time of $t_\mathrm{anneal}=20.4\,\mathrm{ns}$.
This annealing time is very short
compared to the annealing times commonly used by D-Wave quantum annealers (orders of \textmu s).

\begin{table}[th]
  \caption{\textbf{AQA results (success probabilities $P_\mathrm{success}$) obtained by solving the TDSE for Hamiltonians derived from exact cover instances.} Required hardware resources as well as the total run time $t$ are also listed. The annealing scheme is obtained by discretizing the DW\_2000Q\_6 annealing scheme.
  QSDS was used with $n=50$ and $\tau=0.4\unit{ns}$, corresponding to an annealing time $t_\mathrm{anneal}=20.4\, $ns.
  All data was generated on JUWELS-CLUSTER \cite{JUWELS}, except column seven which lists the elapsed times $t_\mathrm{FE}$ that it took four A100 GPUs to solve the exact cover instances by full enumeration.}
  \begin{center}
    \begin{ruledtabular}
      \begin{tabular}{ccccccc}
        instance &    nodes   & processes &  cores    & $t$ [hh:mm]  & $P_\mathrm{success}$  & $t_\mathrm{FE}$ \\
        \hline\noalign{\vskip 4pt}
         30(0) &   64 & 1024  & 3072  & 00:08 & 0.417 & $1.7\, $s  \\
         32(3) &  256 & 4096  & 12288 & 00:14 & 0.237 & $1.7\, $s  \\
         34(3) &  256 & 4096  & 12288 & 00:52 & 0.193 & $2.4\, $s  \\
         36(3) &  256 & 4096  & 12288 & 03:50 & 0.110 & $6.0\, $s  \\
         38(3) &  256 & 4096  & 12288 & 16:40 & 0.085 & $22.3\,$s  \\
         40(0) & 1024 & 16384 & 49152 & 24:40 & 0.038 & $91.8\,$s  \\
       \end{tabular}
    \end{ruledtabular}
    \label{tab:AQA1}
  \end{center}
\end{table}

In Table~\ref{tab:AQA1} we present the data of the AQA simulations with $n=50$ and $\tau=0.4\,\mathrm{ns}$
for exact cover problems with 30, 32, ..., 40 variables.
Column six of Table~\ref{tab:AQA1} shows that the success probability systematically decreases as the number of qubits increases.
This decrease is what one would expect on the basis of the Landau-Zener model
and the assumption that the minimal spectral gap decreases with the system size.
However, AQA uses a time step of $\tau=0.4\,\mathrm{ns}$ that may actually be too large to justify an interpretation in terms of the Landau-Zener model.
Table~\ref{tab:AQA1} also shows that the computational resources required for QSDS to perform these AQA simulations can be considerable.

As already observed earlier~\cite{willsch20_qaoa} and also observed in the AQA simulations, solving the TDSE for model parameters that pertain to D-Wave quantum annealers requires annealing times of the order of nanoseconds to obtain success probabilities of 1\% or better. This observation leads to the conclusion that for the exact cover problems studied here, the annealing time required by TDSE solvers is much shorter than the typical annealing times used by D-Wave quantum annealers, which are of the order of microseconds (see also \cite{Willsch2020FluxQubitsQuantumAnnealing}).
Of course, the TDSE simulations deal with a closed quantum system, free of the interactions with other degrees of freedom which are affecting the operation of real QA devices. Nevertheless, if technically possible, it would be of interest to perform this kind of very fast annealing on genuine quantum annealer hardware.
Finally, it should be mentioned that the wall-clock time required by QSDS (or JUQCS) to cover the nanosecond time span is much larger than a few microseconds, see Table~\ref{tab:AQA1}. Therefore, D-Wave quantum annealers
are very fast simulators in comparison to the software simulators running on conventional semiconductor hardware.

\subsubsection{Comparison of QAOA and AQA}

\begin{figure}[tb]
  \centering
  \includegraphics[width=0.9\columnwidth]{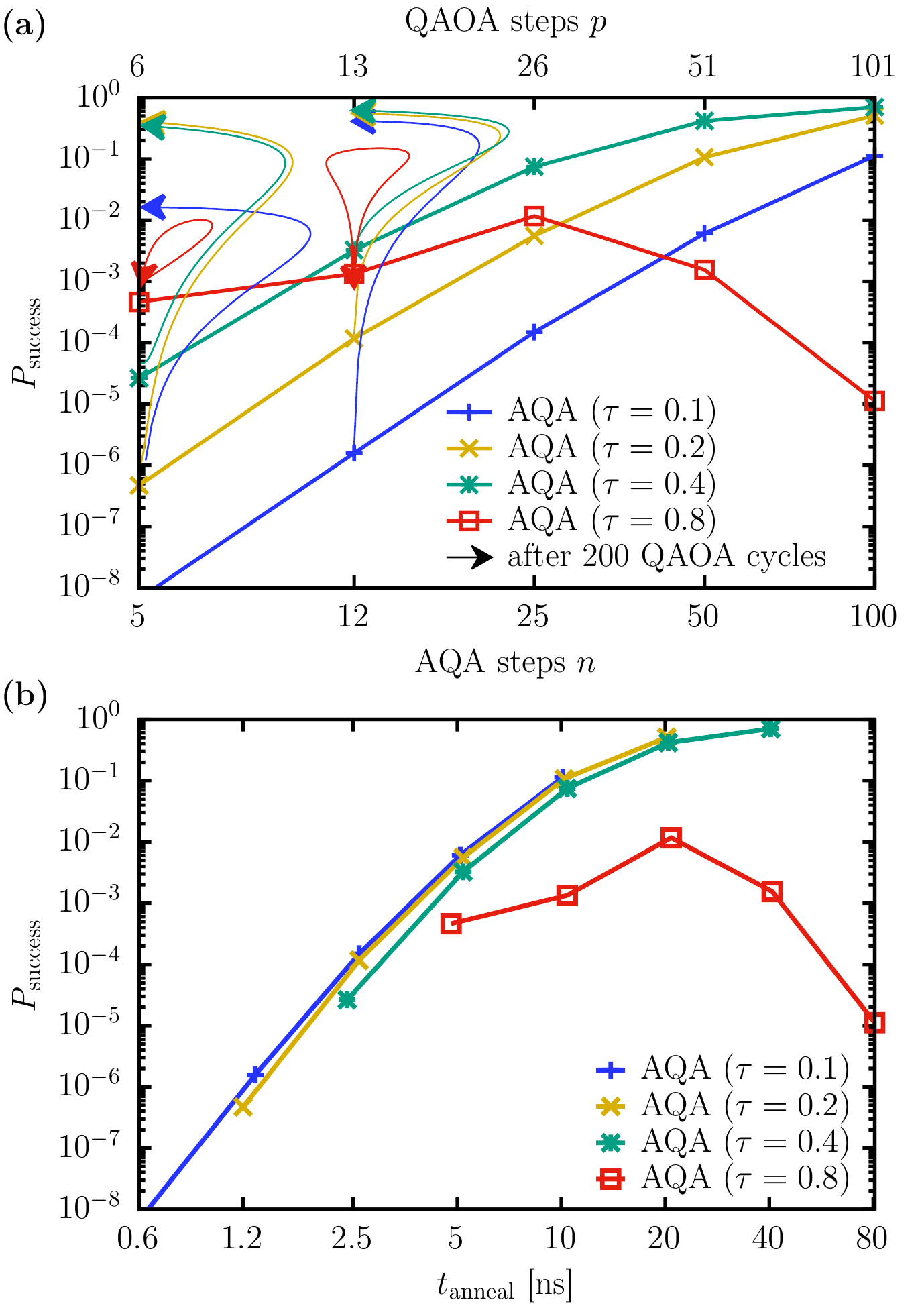}
  \caption{\textbf{Comparison of AQA and QAOA, using exact cover instance 30(0).}
  \textbf{(a)} Success probability as a function of $n$ (AQA, bottom axis) and $p$ (QAOA, top axis).
  In terms of computational effort, AQA with $n$ steps is equivalent to a single JUQCS-call for the QAOA circuit with $p=n+1$, so they are shown together. AQA results are indicated by markers (lines are guides to the eye).
  QAOA results are indicated by the arrows showing the improvement due to the optimization of the variational parameters from the AQA initialization.
  \textbf{(b)} Success probability obtained by AQA as a function of ``annealing time" $t_{\mathrm{anneal}}=(n+1)\tau$ for different values of $\tau$. Lines are guides to the eye.}
  \label{fig:aqa_qaoa_results}
\end{figure}

\begin{figure}
  \centering
  \includegraphics[width=\columnwidth]{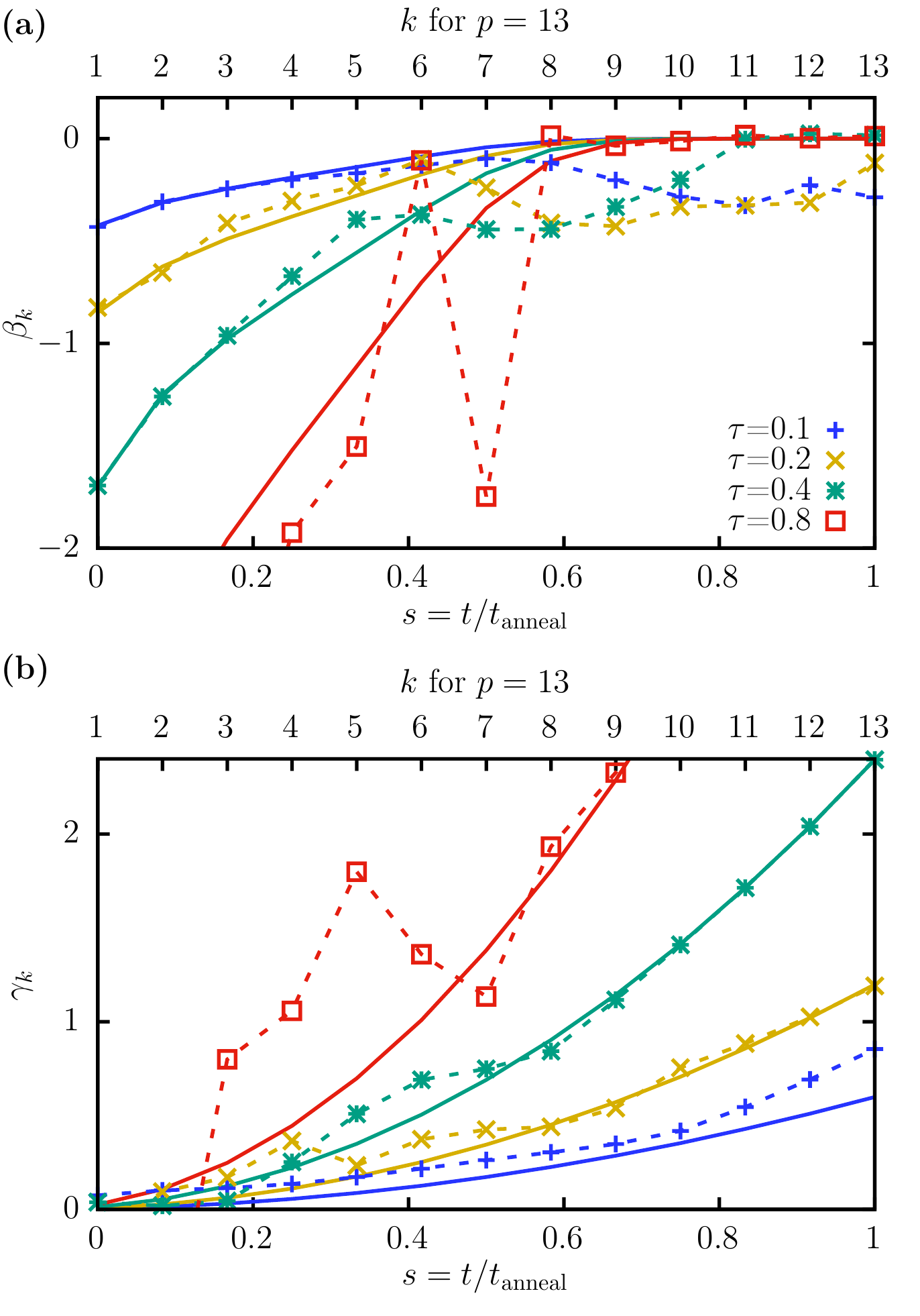}
  \caption{\textbf{Visualization of the variational QAOA parameters (a) $\beta_k$ and (b) $\gamma_k$ for $p=13$ using exact cover instance 30(0).} Solid lines show the initial values taken from the second-order QA initialization (see Fig.~\ref{fig:annealing_schedule} and Eqs.~(\ref{eq:beta_k})--(\ref{eq:gamma_k})), corresponding to the start of the lines with arrows in Fig.~\ref{fig:aqa_qaoa_results}(a). Dashed lines show the final parameters after 200 QAOA optimization cycles (i.e., 200 JUQCS--G calls), corresponding to the end of the lines with arrows in Fig.~\ref{fig:aqa_qaoa_results}(a).
 For $\tau=0.8\,\mathrm{ns}$ (red squares), not the full set of final QAOA parameters is shown to keep the scale reasonable for the other cases and because Fig.~\ref{fig:aqa_qaoa_results}(a) shows that the optimization brings no improvement in the performance.
 Note that the fact that the optimized $\beta_k$ and $\gamma_k$ still roughly follow the initialization from QA suggests that it was an effective modification of the annealing schedule that could so dramatically improve the success probability in Fig.~\ref{fig:aqa_qaoa_results}(a).}
  \label{fig:optimized_annealing_schedule}
\end{figure}

Results to compare QAOA and AQA are presented in Fig.~\ref{fig:aqa_qaoa_results}(a). It shows the success probability as a function of the number of AQA steps $5\le n\le100$ for different values of $\tau$. Additionally, the arrows for $p=6$ and $p=13$ show the results obtained after optimizing the corresponding $\beta$ and $\gamma$ with the QAOA (using SLSQP) after 200 JUQCS--G calls. The initial and final values for  $\beta$ and $\gamma$ are shown in Fig.~\ref{fig:optimized_annealing_schedule}. 

We compare QAOA and AQA in terms of computational work. Performing QAOA with $p$ steps and $m$ optimization cycles (i.e., $m$ calls to JUQCS--G) needs computational work proportional to $m\times p$. Performing AQA with $n$ is equivalent to performing QAOA with $p=n+1$ and $m=1$ (as AQA only needs a single call to JUQCS--G), so the computational work equivalent for AQA is $n+1$. Thus, for QAOA to compete with AQA, it should use $m<(n+1)/p$ optimization cycles to reach a similar success probability.

However, as already mentioned and seen in Table~\ref{tab:qaoa_results}, the number of JUQCS--G calls $m$ required to obtain high success probabilities with the QAOA is fairly large. And as Fig.~\ref{fig:aqa_qaoa_results}(a) shows, sometimes even $m=200$ optimization cycles (with computational work $200p$) are not enough to reach the success probabilities that AQA reaches already after $n=100$ steps.
Hence, for the exact cover instances considered, QAOA cannot compete with AQA in terms of computational efficiency.

For AQA, we find that the success probability increases for increasing number of steps $n$.
As Fig.~\ref{fig:aqa_qaoa_results}(b) shows, the main increase in the success probability is due to the increased annealing time.
The success probability also increases with $\tau$, up to a certain point where no further improvement is made.
For $\tau=0.8\unit{ns}$, we find that for a fixed annealing time the probability is substantially lower than for the other values of $\tau$.
Here, the time step $\tau=0.8\unit{ns}$ is too large to justify even a crude approximation of an annealing schedule.

We believe that AQA is best seen as a viable heuristic, requiring a few numerical experiments to optimize the parameters (in contrast to the QAOA which usually needs many iterations to obtain a reasonable result). To some extent, Fig.~\ref{fig:aqa_qaoa_results}(b) gives a hint for an explanation why AQA works well. For small time steps ($\tau=0.1$), AQA  is essentially the same as slow quantum annealing which, according to the adiabatic theorem, should give the ground state of the problem Hamiltonian. If the time step is too large ($\tau=0.8$), we lose contact with the idea of quantum annealing. We still get reasonable success probabilities for $t<20$, but if we then anneal longer, the success probability drops. We have started a new project that specifically studies AQA to address this aspect.

The fact that QAOA is able to optimize the cases $\tau\in\{0.1,0.2,0.4\}\unit{ns}$ can be interpreted as follows: For $\tau$ up to $0.4\unit{ns}$, AQA still resembles QA with a very short annealing time (e.g.~by rendering the system in a low energy state as in~\cite{Hsu2019AnnealPathFloppiness}), so optimization can increase the success probability (as indicated by the arrows in Fig.~\ref{fig:aqa_qaoa_results}(a)).
The case $\tau=0.8\unit{ns}$, however, does not seem to yield suitable initial values for the parameters $\beta_k$ and $\gamma_k$ as is clear from the fact that the optimization during QAOA does not yield a significant improvement.
However, for AQA with a small number of steps $n$ (e.g.~$n=5$ where $P_\mathrm{success}\approx10^{-3}$ in Fig.~\ref{fig:aqa_qaoa_results}(a)), it may still be a reasonable choice (see also the surprisingly good scaling in Fig.~\ref{fig:aqa_qaoa_scaling} below).

On the one hand, we find that with AQA for a large number of steps ($n\approx 50$--$100$), we obtain similar success probabilities as with QAOA for smaller $p\approx 6$--$13$, but the QAOA optimization requires many calls to JUQCS--G.
Moreover, we also observed that the minimizer can get stuck in a local optimum which then does not lead to an improved performance over AQA even for the same number of steps and many more circuit evaluations.
However, also for AQA, we have to search for a good value of $\tau$ which optimizes the success probability for a given number of steps $n$. The same $\tau$ that leads to an optimal success probability for a certain value of $n$ may not be optimal for other values of $n$.
Still, for AQA, we basically have to optimize a single parameter only (if $n$ is fixed) and not $2p$ parameters as is the case for QAOA.

On the other hand, for NISQ devices, AQA with a large number of steps $n$ (and equivalently QAOA with a large number of steps $p$) will probably suffer from accumulated errors during the relatively long quantum circuit. Thus, NISQ devices may cope better with QAOA with small $p$ than AQA with large $n$. Perhaps, building on the result that the optimized $\beta_k$ and $\gamma_k$ in Fig.~\ref{fig:optimized_annealing_schedule} were not far from the annealing initialization, the best solution might be to indeed use AQA with small $n$ but with better effective (maybe problem-dependent) annealing schedules. Comparing the performance of AQA and QAOA on NISQ devices in practice would be an interesting study for the future.

\subsubsection{Scaling as a function of the problem size \texorpdfstring{$N$}{N}}

\begin{figure}[tb]
  \centering
  \includegraphics[width=\columnwidth]{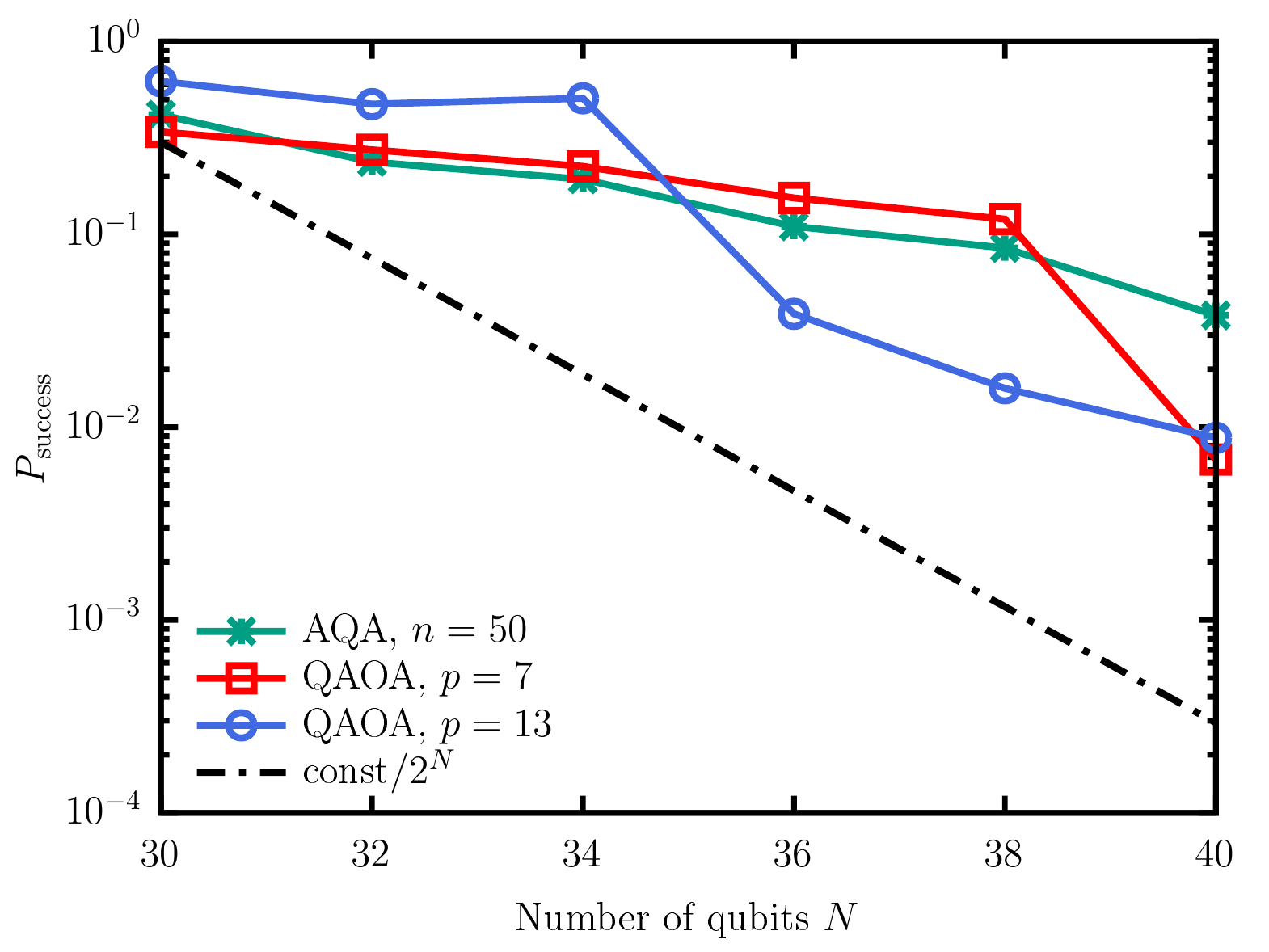}
  \caption{\textbf{Scaling of the success probability as a function of the system size $N$}. The different markers correspond to AQA with $n=50$ and $\tau=0.4\unit{ns}$ (green asterisks), QAOA for $p=7$ (red squares), and QAOA for $p=13$ (blue circles), taken from Tables~\ref{tab:qaoa_results} and \ref{tab:AQA1}. The dash-dotted line indicates the scaling of a uniform probability distribution. The green asterisk at $N=30$ is the same point shown in Fig.~\ref{fig:aqa_qaoa_results} at $n=50$, $\tau=0.4\unit{ns}$ and $t_{\mathrm{anneal}}=20.4\,\mathrm{ns}$. Lines are guides to the eye.}
  \label{fig:aqa_probabilities}
\end{figure}

\begin{figure}[tb]
  \centering
  \includegraphics[width=\columnwidth]{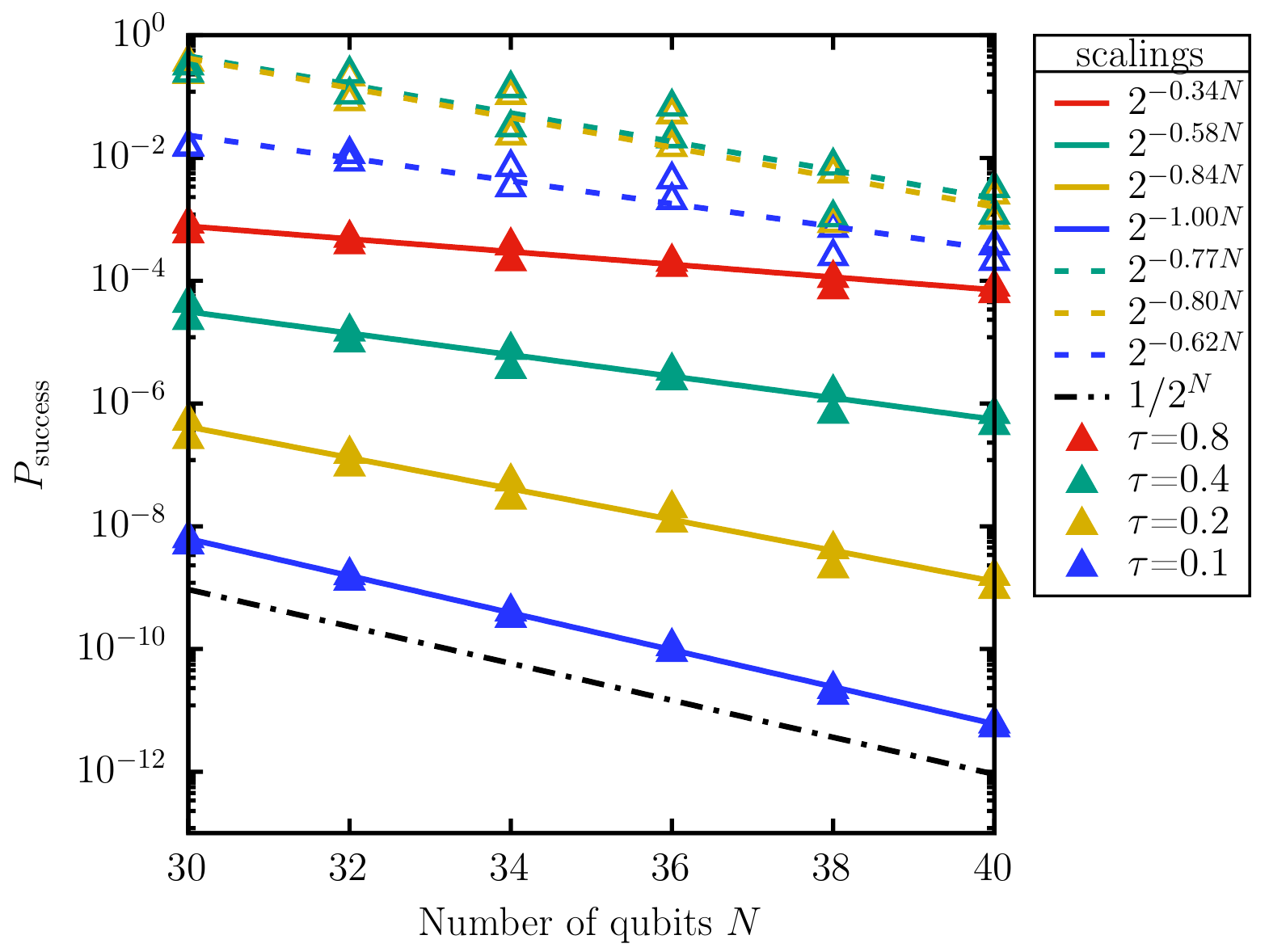}
  \caption{\textbf{Scaling of the success probability as a function of the system size $N$, using AQA with $n=5$ (filled triangles) and pre-optimized QAOA with $p=6$ (open triangles).} Here, pre-optimized means that for all instances, the same values for $\beta_k$ and $\gamma_k$ (obtained from the optimization of the 30-qubit problem instance 30(0), see Fig.~\ref{fig:aqa_qaoa_results}) are used.
  We ran two problem instances for each system size, so all triangles appear in pairs. Solid (dashed) lines show fits to the AQA (QAOA) results.
  Different colors correspond to different values for $\tau$ as indicated in Fig.~\ref{fig:aqa_qaoa_results}. The dash-dotted line indicates the probability to find the solution when picking from a uniform distribution at random. The data for all runs was obtained with JUQCS--G, using a quantum circuit that performs the time evolution simulated by QSDS (see \equref{AQA4}), thereby leveraging the computational power of the GPUs.}
  \label{fig:aqa_qaoa_scaling}
\end{figure}

In Figs.~\ref{fig:aqa_probabilities} and~\ref{fig:aqa_qaoa_scaling}, we show the scaling of the success probabilities obtained for different problem instances with increasing number of qubits using AQA and QAOA.
In Fig.~\ref{fig:aqa_probabilities}, the scalings of AQA and QAOA with the system size look quite similar up to $N=34$. For larger $N$, the drops in the success probability for the QAOA data are where the minimizer probably got stuck in a local optimum (red entries in Table~\ref{tab:qaoa_results}). We note that for QAOA, we ran the optimization procedure for each system size. For AQA, we did not perform any optimization but we used a relatively large step size $\tau$.

In Fig.~\ref{fig:aqa_qaoa_scaling}, we always use the same $\beta_k$ and $\gamma_k$ obtained from the QAOA optimization for problem instance 30(0). In other words, we take the variational parameters obtained by optimizing the 30-qubit instance 30(0), and we use the \emph{same} parameters for \emph{different} problem instances of \emph{different} size. In this way, we test how well the effective ``modified annealing schedule'' (cf.~Fig.~\ref{fig:optimized_annealing_schedule}) generalizes to other problems of larger size.

Figure~\ref{fig:aqa_qaoa_scaling} shows that the QAOA parameters generalize systematically, but as expected, the success probability still drops exponentially with increasing qubit number. Furthermore, the fits to the data (lines) show that the exponential scaling is of the form $2^{-\alpha N}$ for $\alpha\gtrsim0.6$. In contrast, we observe that for AQA, although the drop in success probability is also exponential, the exponent $\alpha$ behaves more favorably. Remarkably, this favorable scaling is especially pronounced for the large value of $\tau=0.8\,\mathrm{ns}$ (where $\alpha=0.34$), which is very far in the AQA regime. 

\section{Summary}
\label{sec:conclusion}

The first part of this paper was devoted to the study of the weak and strong scaling behavior of a GPU-accelerated version (JUQCS--G) of  the  J\"ulich  Universal  Quantum Computer Simulator (JUQCS)~\cite{deraedt18} by performing
benchmarks on JUWELS Booster, a supercomputer with 3744 NVIDIA A100 Tensor Core GPUs.
Our data shows that JUQCS--G exhibits nearly perfect weak and strong scaling for systems up to 42 qubits.
Comparing elapsed times for JUQCS--G and for JUQCS--E, a non-GPU version of JUQCS, shows that the former is a factor of 10--18 faster than the latter. As the number of qubits reaches the maximum that the available memory allows, the larger fraction of the elapsed time goes into MPI communication, for both the GPU and non-GPU version. In any case, using the GPU version significantly reduces the computing time required to simulate quantum computers and quantum systems. 

In the second part of the paper, we have used JUQCS--G to solve exact cover problems with up to 40 variables (qubits).
Hereby the focus was on the assessment of the potential of the quantum approximate optimization algorithm (QAOA) as a vehicle to solve optimization problems involving 30--40 qubits.
Due to the minimization of parameters reflecting the variational nature of the QAOA, it is necessary to execute the quantum circuit many times.
In most cases, at least for the 30--40 qubit instances that we have studied, the number of repetitions (with different sets of parameters) has a negative impact on the efficiency of the QAOA. 

As an alternative, we also studied the performance of what we called approximate quantum annealing (AQA). AQA is a discretized version of quantum annealing which is approximate in the sense that we use only a few, relatively large time steps,
possibly beyond the regime where quantum annealing is theoretically justified through the adiabatic theorem.
Nevertheless, we found that, without any optimization, we already obtain success probabilities $\gg 1\%$ for problem instances up to $N=40$ qubits.
These promising results suggest that for future gate-based quantum computers which can cope with a larger circuit depth, direct AQA may provide a better alternative to the QAOA as it avoids the costly optimization procedure.
As a matter of fact, from a computational viewpoint, AQA is much more efficient than the QAOA.

It is self-evident that all the simulation results that we have presented in this paper have been obtained by simulating the ideal mathematical model of a gate-based quantum computer.
In this sense, the 30--40 qubit results presented in this paper are the ``best case'',
very unlikely to be achieved by using a real quantum processor.
Of course, it is possible to incorporate noise and errors into our simulations (left for future work), but accounting for the intrinsic quantum gate errors of 30--40 qubit systems requires simulation times that are currently prohibitive~\cite{WillschDennis2020Phd}.
Clearly, to get a view on the errors involved, it would be very interesting to run say a 30-qubit exact cover quantum circuit on a NISQ device and compare the experimental data with the simulation results.
Furthermore, as our conclusions are drawn from results obtained for 30--40 variable exact cover problems, it might be of interest to investigate how generic these conclusions are by studying different types of optimization problems. 

\begin{acknowledgments}
The authors gratefully acknowledge the Gauss Centre for Supercomputing e.V.
(www.gauss-centre.eu) for funding this project by providing computing time
on the GCS Supercomputer JUWELS at J\"ulich Supercomputing Centre (JSC).
We would like to thank M. Svensson for providing the exact cover problem instances.
We gratefully acknowledge support during the JUWELS Booster Early Access period by A. Herten, M. Hrywniak, J. Kraus, A. Koehler, P. Messmer, M. Knobloch as well as the JUWELS Booster Project Team (JSC, Atos, ParTec, NVIDIA).
D.W.'s work was partially supported by the Q(AI)$^{2}$ project. 
D.W. and M.W. acknowledge support from the project J\"ulich UNified Infrastructure for Quantum computing (JUNIQ) that has received funding from the German Federal Ministry of Education and Research (BMBF) and the Ministry of Culture and Science of the State of North Rhine-Westphalia. 
Open Access publication was funded by the Deutsche Forschungsgemeinschaft (DFG, German Research Foundation, grant number 491111487).
\end{acknowledgments}

\begin{widetext}
\appendix

\section{Second-order initialization of the QAOA parameters}
\label{app:initialization}
The QAOA state with $2p$ variational parameters $\beta_k$ and $\gamma_k$ reads (see \equref{eq:QAOA_state})
\begin{align}
    \ket{{\beta,\gamma}} = e^{-i\beta_pH_D}e^{-i\gamma_pH_C}\cdots e^{-i\beta_1H_D}e^{-i\gamma_1H_C}\ket{+}^{\otimes N}.
\end{align}
Inserting the values for the QAOA parameters given in Eqs.~(\ref{eq:beta_k})--(\ref{eq:gamma_k}), and replacing $\ket{+}^{\otimes N}$ by $e^{i\tau A(s_1)H_D/2}\ket{+}^{\otimes N}$ (which only differs from $\ket{+}^{\otimes N}$ by a global phase and is thus physically equivalent) yields
\begin{align}
    \ket{{\beta,\gamma}} = e^{i\tau A(s_p)H_D/2}e^{-i\tau B(s_p)H_C}\cdots e^{i\tau(A(s_2)+A(s_1))H_D/2}e^{-i\tau B(s_1)H_C}e^{i\tau A(s_1)H_D/2}\ket{+}^{\otimes N}.
\end{align}
Here we see that $\ket{{\beta,\gamma}}$ can be expressed as
\begin{align}
    \ket{{\beta,\gamma}} = U(s_p)\cdots U(s_1)\ket{+}^{\otimes N},
\end{align}
where $U(s_k)$ for $k=1,\ldots,p$ is the second-order Suzuki-Trotter decomposition \cite{DeRaedt83GeneralizedTrotter,Suzuki1985ProductFormulaError},
\begin{align}
    U(s_k) = e^{i\tau A(s_k)H_D/2} e^{-i\tau B(s_k)H_C} e^{i\tau A(s_k)H_D/2},
\end{align}
of the discretized time-evolution operator generated by the QA Hamiltonian $H(s) = A(s)(-H_D) + B(s)H_C$ (see \equref{eq:Hofs}). We note that besides the choice $s_k=(k-1)/(p-1)$ taken in this paper, also the mid-point decomposition used in \cite{willsch20_qaoa} is a good choice for the discretization (cf.~\cite{Suzuki1993GeneralDecompositionTheoryOrderedExponentials}).
\end{widetext}

\bibliographystyle{apsrev4-2custom}
\bibliography{bibliography}

\end{document}